\documentclass[sort&compress,12pt]{elsarticle}
\usepackage{graphicx}  
\usepackage{amsmath}
\usepackage{amssymb}
\usepackage{epstopdf}
\usepackage{subfigure}
\usepackage{setspace}
\usepackage{latexsym}
\usepackage{amsfonts}
\usepackage{color}
\usepackage{ctable}
\usepackage{multirow}
\usepackage{natbib}
\usepackage{url}

\journal{Computer Networks}

\begin{document}
\newcommand{\GT}{\mathcal{G}}
\newcommand{\PT}{\mathcal{P}}
\newcommand{\CT}{\mathcal{C}}
\newcommand{\ET}{\mathcal{E}}
\newcommand{\VT}{\mathcal{V}}
\newcommand{\figs}{.}

\newcommand{\TB}[1]{\textcolor{black}{#1}}
\newcommand{\rk}[1]{{[\textcolor{black}{#1}]}}
\newcommand{\BM}[1]{{\textcolor{black}{#1}}}

\newcommand{\marginnote}[1]{\marginpar{{\parbox{\linewidth}{\hrule\setstretch{.05}\textcolor{blue}{\scriptsize \sf #1}}}}}

\begin{frontmatter}

\title{Intrinsically Dynamic Network Communities}
\tnotetext[label1]{All authors contributed equally to the manuscript.}

\author[ISCPIF,CREA]{Bivas Mitra}
\author[LIP6,CREA]{Lionel Tabourier}
\author[ISCPIF,CAMS,CREA]{Camille Roth}

\address[ISCPIF]{Institut des Syst\`emes Complexes-Paris-Ile-de-France, 57 rue Lhomond, F-75005 Paris}
\address[LIP6]{LIP6, Universit\'e Pierre et Marie Curie, 4 place Jussieu, F-75005 Paris}
\address[CAMS]{CAMS, CNRS/EHESS, 190 avenue de France, F-75013 Paris}
\address[CREA]{CREA, CNRS/Ecole Polytechnique, 42 bd Victor, F-75015 Paris}

\begin{abstract}
Community finding algorithms for networks have recently been extended to dynamic data.  Most of these recent methods aim at exhibiting community partitions from successive graph snapshots and thereafter connecting or smoothing these partitions using clever time-dependent features and sampling techniques.  These approaches are nonetheless achieving \emph{longitudinal} rather than \emph{dynamic} community detection.  We assume that communities are fundamentally defined by the repetition of interactions among a set of nodes over time.  According to this definition, analyzing the data by considering successive snapshots induces a significant loss of information: we suggest that it blurs essentially dynamic phenomena --- such as communities based on repeated inter-temporal interactions, nodes switching from a community to another across time, or the possibility that a community survives while its members are being integrally replaced over a longer time period.  We propose a formalism which aims at tackling this issue in the context of time-directed datasets (such as citation networks), and present several illustrations on both empirical and synthetic dynamic networks.  We eventually introduce intrinsically dynamic metrics to qualify temporal community structure and emphasize their possible role as an estimator of the quality of the community detection --- taking into account the fact that various empirical contexts may call for distinct `community' definitions and detection criteria.
\end{abstract}

\begin{keyword}
community detection \sep dynamic networks \sep citation networks \sep community quality metrics

\end{keyword}

\pagestyle {plain}

\end{frontmatter}

\section{Introduction}

The detection of groups from structural interaction datasets has generated a tremendous amount of literature over the past decade \cite[for a recent and comprehensive review, see][]{fort:comm}.
These methods essentially aim at partitioning nodes into ``communities'' in such a way that the density of links within communities is higher than the density of links between them -- thereby adopting one possible structural definition of social communities, that ``its members should have many relations with each other and few with non-members''  \cite[p. 121]{alba:grap}.
Most corresponding studies are primarily featuring static community finding algorithms, \hbox{i.e.} relying on networks as static graphs, either derived from aggregation of data over the whole crawling period or from a snapshot at a particular point in time --- which is essentially the same.

Emphasizing the fact that real-world communities are also time-evolving, a more recent stream of research endeavored at describing the dynamics of such communities \cite{hoft,oliv,oliv1,greene,spiliopoulou2006monic,asur,falk,toyoda,palla,gul,chak,chi,lin,kim,jdidia,pons,mucha}.
In this portion of the literature, communities and their evolution are however being studied in a somewhat independent manner: communities are separately extracted from several snapshots and then time-dependent characteristics are introduced to smooth potential differences between the various partitions over time, or to connect various partitions in order to have inter-temporally linked communities.  In other words, these approaches are rather proposing a longitudinal community detection.

Such methodology may however become inappropriate if some intricate temporal relationships occur, in particular if some nodes which were part of a community at some point are not anymore at a later point in time, while the ``underlying group'' remains --- \hbox{e.g.} an association, a lab, a company which survives through time without depending on a specific set of permanent members.\footnote{This problem was originally framed by G. Simmel in the following qualitative terms: ``\emph{The most general case in which the persistence of the group presents itself as a problem occurs in the fact that, in spite of the departure and the change of members, the group remains identical. We say that it is the same state, the same association, the same army, which now exists that existed so and so many decades or centuries ago. This, although no single member of the original organization remains.}'' \cite[p.~667]{simm:pers}}
More broadly, our first assumption is that communities are fundamentally defined by a certain amount of interaction recurrence among a possibly disparate set of nodes over time.  This type of phenomenon is certainly roughly detectable through the trace of interactions, as recorded in a static graph, or a series of (snapshot-based) static graphs.  We suggest however that the aggregation operation may induce a strong information loss and often requires additional thresholding hypotheses (on link weight and timewindow width, \emph{inter alia}). Our approach aims at remaining as much as possible faithful to the original empirical data.

\bigskip We therefore propose a simple formalism to detect essentially dynamic communities from a \emph{diachronic dataset}, and understand their evolution within a unified, single-pass procedure. By ``diachronic dataset'' we mean that the data we consider features links whose extremities are associated with possibly distinct timestamps, like in a citation network or an email network (an item of B made at $t'$ is cited/answered by A at $t>t'$).

After a presentation of the relevant literature in Sec.~\ref{sec:relatedwork}, we explain (Sec.~\ref{sec:methodology})  the principles of this method which consists in a new ontological viewpoint built upon usual concepts of community detection methods. In Sec.~\ref{sec:datasources} we introduce various datasets, both synthetic and empirical (citation) networks, in blog- and science-related contexts. We then discuss results in Sec.~\ref{sec:results} and, in particular, we propose a variety of metrics aimed at providing information about the type and quality of detected temporal communities. This leads to a macro-level characterization of empirical datasets through their dynamic community profile. Focusing on the behavior of the individual nodes within communities --- for instance how many times nodes appear or change their community membership --- we later present in Sec.~\ref{sec:behavior} several features of temporal communities which are typical of our approach.  We discuss eventually (Sec.~\ref{sec:discussion}) how this approach exhibits novel stylized facts that are not directly obtainable from existing methods, and how our methodology could be applied, with some tweaking, to other kind of network data where edges are associated with discrete timesteps. In particular, we also suggest a possible improvement of our methodology by introducing an additional ``temporal-oriented'' step which can suitably complement the `structure-oriented' approach of our algorithm.

\section{Related work}\label{sec:relatedwork}

The analysis of community evolution in dynamic networks is a recent and increasingly active field.
To our knowledge, the first study was carried out by Hopcroft et al.~\cite{hoft} who analyzed several snapshots of a scientific citation database. Using a hierarchical clustering algorithm --- based on the similarity of articles --- they defined ``natural communities'' particularly robust to minor perturbations of the graph. Then they proposed rules based on set theory tools to decide how communities evolve from a snapshot to another. Many variations of this set of rules have been proposed afterwards \cite{oliv,oliv1,greene}, one of the most comprehensive being {\sc monic} \cite{spiliopoulou2006monic}.
Since then, such approaches have been widely used and developed.
Asur et al.~\cite{asur} introduced a family of events (merging, splitting etc.) on both communities and individuals to characterize the evolution. Falkowski et al.~\cite{falk} computed groups with a classical divisive algorithm at each timestep,
links were created between community instances at different timesteps with a weight depending on {\sc monic} matching rules.
Toyoda and Kitsuregawa~\cite{toyoda} studied the evolution of Web communities from a sequence of snapshots. Palla et al.~\cite{palla} used the clique percolation method to extract communities at each moment and then matched them at consecutive time steps to make statistics on their evolution: sizes, ages, overlaps between two different time steps etc.
In~\cite{gul} three classical algorithms are used to detect communities at every snapshots, and since these methods are unstable, a stabilization of the Louvain algorithm is proposed, achieving simultaneously high modularity and good stability.


The above-described approaches are basically two-stage methods: (i) detect clusters independently at some period in the graph evolution, then (ii) infer relationships between partitions at different periods.
However, this kind of approach is susceptible to edge effects, due to the longitudinal analysis: discrete time periods have to be defined and they sometimes split important signals into two distinct periods. Significant variations even between partitions close in time may appear, which results in artifactual community structure evolution.
It is desirable, instead, to have a unified framework, in which cluster evolution can be deduced both from the current graph structure and from the knowledge of the community structure at previous time steps.
More recently, a new kind of clustering concept called ``evolutionary clustering'' has been proposed to capture the evolutionary process of clusters in temporal data.
This framework was introduced by Chakrabarti et al.~\cite{chak} who assume that short-term community structure alterations are not desirable and therefore smooth communities over time. The smoothing process consists of a trade-off between the quality of the detection on a snapshot and its consistency with the previous community detection.
Chi et al.~\cite{chi} extended similar ideas and proposed the first evolutionary spectral clustering algorithm.
Based on evolutionary clustering, Lin et al.~\cite{lin} introduced the framework {\sc FacetNet}, which allows vertices to belong to several communities simultaneously.
{\sc FacetNet} can be extended to handle vertex addition and removal as well as variations of the number of clusters in consecutive timesteps.
However, it is not able to account for the creation and disintegration of communities,
and it is not scalable to large systems because of computational limits.
These issues have been addressed in a recent approach by Kim and Han~\cite{kim}.
A uniform community detection methodology called OSLOM (Order Statistics Local Optimization Method) has been
recently proposed in~\cite{plosone}. 
It detects clusters in networks accounting for edge directions, edge weights, overlapping communities, hierarchies and community
dynamics. The algorithm is based on the local optimization of a fitness function expressing the statistical significance of clusters with
respect to random fluctuations, which is estimated with tools of Extreme and Order Statistics~\cite{lancichinetti-2010-81, radicchi-2010-82}. Instead of separately analyzing the snapshots, this methodology aims at combining information from different time slices, 
thereby taking advantage of the information from different snapshots to uncover correlations between structures of the system at various time stamps.


The temporal information can also be integrated in the graph itself.
One attempt in this direction is~\cite{jdidia} which uses the community detection algorithm Walktrap~\cite{pons} but changes the input graph to integrate dynamics: they create a temporal graph by duplicating nodes appearing at different timesteps, and by artificially linking a node to itself at the following timestamp. Communities are then detected on this temporal graph which contains a mix of temporal and classical links. This idea of connecting several snapshots is extended in~\cite{mucha}:
the authors propose a more general way to build the temporal graph and suggest an appropriate modification of the modularity.
Yet these approaches rely on the arbitrary hypothesis that a connection between two nodes can be compared to the connection of a node with itself across time.

The formalism we propose shares some features with these temporal graphs, but avoids ad hoc alterations of the underlying empirical data, to the cost of being restricted --- at least in this paper ---  to a certain kind of data where link extremities are explicitly attributed possibly different timestamps, as is the case \hbox{e.g.} in citation networks.


\section{Methodology}\label{sec:method}\label{sec:methodology}

Our methodology explicitly relies on a diachronic dataset, \hbox{i.e.} a set of links connecting a timestamped source to a timestamped destination, where source and destination timestamps may be distinct.  This is typically the case in citation networks.\footnote{See Sec.~\ref{sec:discussion} for a discussion on possible extensions of this method.}
We define basic notations to represent such citation datasets and describe a graph suited to this kind of data.


\subsection{Physical graph vs. temporal graph}


We call \emph{physical nodes} the source or destination of the links -- without consideration for timestamps --
let $V =\{v_i\}_{i\in\{1,..,n\}}$ be the associated set.
The timescale is supposed to be discretized so that any timestamp is an element of set $T=\{t_i\}_{i\in I}$.

The empirical data consists of directed links whose extremeties relate to possibly distinct timesteps:
\begin{equation}
\Psi=\Big\{\Big((v_i,t_i),(v_j,t_j)\Big) \mid v_i,v_j\in V \ , \ t_i,t_j\in T\Big\}
\end{equation}

Typical approaches aim at aggregating this original data over the whole timescale, thereby obtaining a usual directed graph $G(V,E)$ whose edge set is $E=\{(v_i,v_j)\}$, where node $v_i$ cites node $v_j$. We will refer to such links as \emph{physical links} and to $G$ as the \emph{physical graph}.

To our knowledge, most dynamic community detection methods are relying on this type of graphs: they ultimately rely on physical graphs based on a particular slicing of the dataset into time periods (usually, a partition of time periods). Our point is to diverge from this significant information reduction by relying directly on the original empirical data.

\bigskip

\begin{figure}
\centering\includegraphics[width=.7\linewidth]{\figs/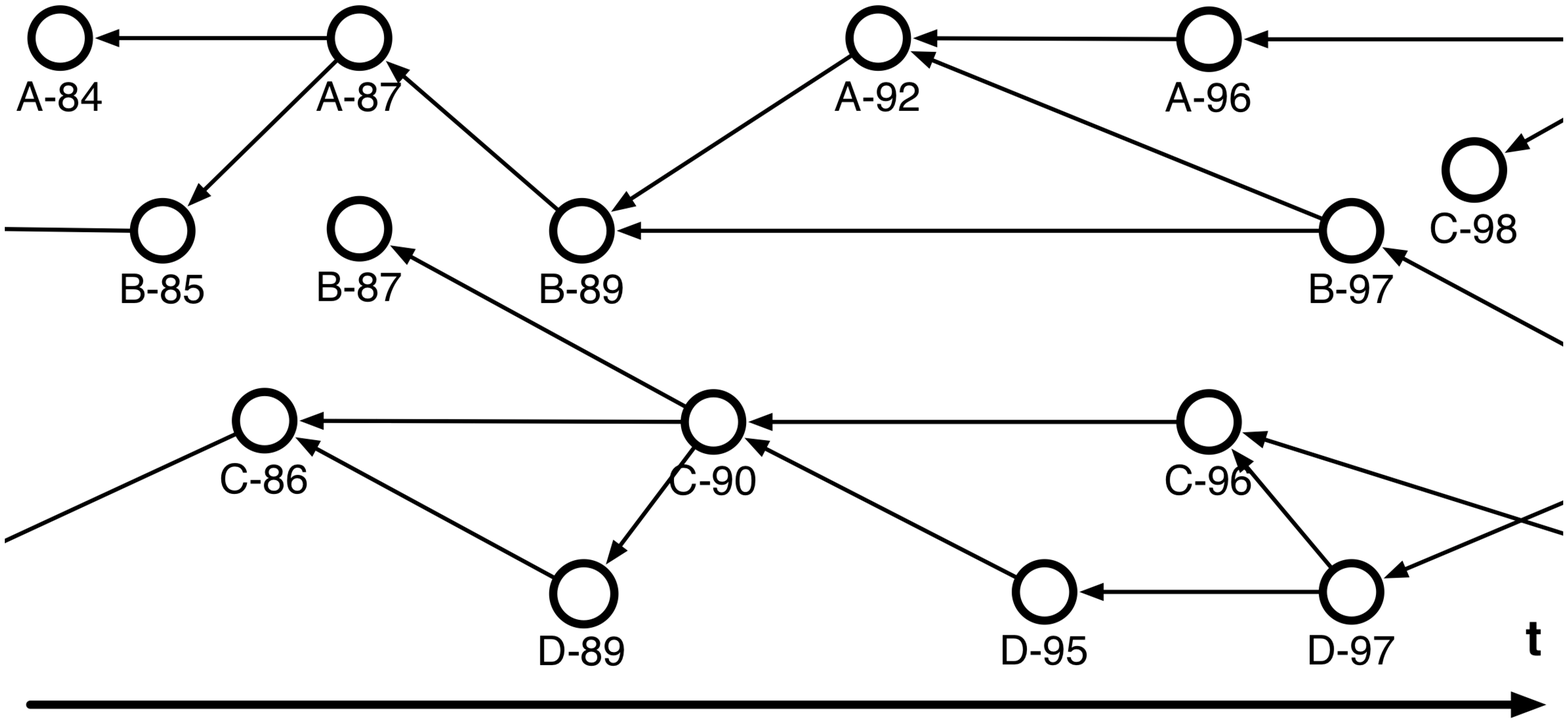}
\caption{\label{fig:temporal}Example of a temporal graph based on four physical nodes A, B, C and D.}
\end{figure}

We therefore define a key concept for the following analysis:
the \emph{temporal graph} $\GT=(\VT,\ET)$, defined upon a set of time-labeled vertices, or \emph{temporal nodes}, $\VT=\{(v_i,t_i)\}$, and a set of time-labeled links, or \emph{temporal links}, $\ET=\Big\{ \Big(  (v_i,t_i),(v_j,t_j) \Big)\Big\}$, such that $\forall i , (v_i,t_i) \in \VT$.
Practically, the $t_i$ and $t_j$ associated to $v_i$ and $v_j$ will be elements of $T$ corresponding to the timestamp when $v_i$ cites $v_j$, as defined in $\Psi$. See an illustration on Fig.~\ref{fig:temporal}.

\subsection{Temporal community detection}

Contrary to the physical graph $G(V,E)$, the temporal graph $\GT(\VT,\ET)$ built from $\Psi$ provides us with information regarding the appearance as well as disappearance of citation links between specific physical nodes over a period of time.
We may now apply a classical community detection algorithm on $\GT$ to obtain a cover $\CT=\{C_1, C_2,....,C_m\}$ of the temporal node set $\VT$. 

Each of these $C_i$ is a set of temporal nodes from the temporal graph. Combined with $\GT$, and more precisely $\ET$, each $C_i$ straightforwardly and uniquely defines a subgraph induced by the corresponding subset of $\VT$. It will be called a \emph{temporal community}.

\bigskip
In the following practical explorations, we rely on the so-called Louvain algorithm \cite{blon:fast} --- essentially for its ability to quickly cluster temporal nodes into cohesive subgraphs. $\CT$ will be a partition because this very algorithm does output partitions of the underlying graph nodes, as do most network algorithms for community detection.  Here we would like to emphasize the fact that our methodology is usable with other typical static community finding algorithms, including overlapping community detection techniques \cite[e.g.,][]{evan:cliq}. As an evidence, in the next section we illustrate our methodology with two other algorithms such as Girvan-Newman~\cite{grivannewman} and Walktrap~\cite{pons} in addition to Louvain algorithm for the synthetic dataset (Fig.~\ref{fig:variousalgo}).


\section{Data sources}\label{sec:datasources}

\subsection{Synthetic data}\label{sec:exp}

We implement in this section our methodology on a synthetically generated citation data set, built using an a priori community structure.  This synthetic data will be useful as a toy example to demonstrate the effect of the method on a controlled environment, where the level and extent of recurrent, dynamic interaction is \emph{a priori} known.

\paragraph{Synthetic data generation}\label{sec:syn_data}
To produce the synthetic dataset, we first consider $n_c$ a priori ``communities''. We create a set of $n = m\cdot n_c$ physical nodes $V=\{v_1,...,v_n\}$ each assigned  a community number in such a way that each community has a constant number of $m$ members.
\newcommand{\tmax}{t_{\text{max}}}
We also define a set of timestamps $T=\{t_i\}_{i\in\{1,...,\tmax\}}$. We thus have a 2D-grid of size $n \cdot \tmax$ whose elements are temporal nodes of $\VT$. We eventually fix a temporal graph density by tuning the average out-degree of the temporal nodes $d$.

We then synthesize a dynamic citation network $\GT^p$ by randomly generating a number of temporal links respecting the following constraints:
\begin{itemize}
\item A link from $(v,t)$ to $(v',t')$ must be such that $t'\leq t$.
\item The total number of outgoing temporal links emanating from all temporal nodes at a specific timestamp $t$ is given as $d\cdot n$. Outgoing links are spread randomly on the $n$ nodes at each timestep $t$.
\item An outgoing temporal link has a probability $p$ to reach a temporal node of the same community as the source and probability $1-p$ to reach a temporal node assigned to any other \emph{a priori} community.
\item We additionally introduce a sliding window of size $w$ so that $t-w \leq t'$, ensuring that most sources refer to recent destinations rather than to arbitrarily old ones.
\end{itemize}

\begin{figure*}[h!]
\centering
\subfigure[$p=1$]{\includegraphics[width=.31\linewidth]{\figs/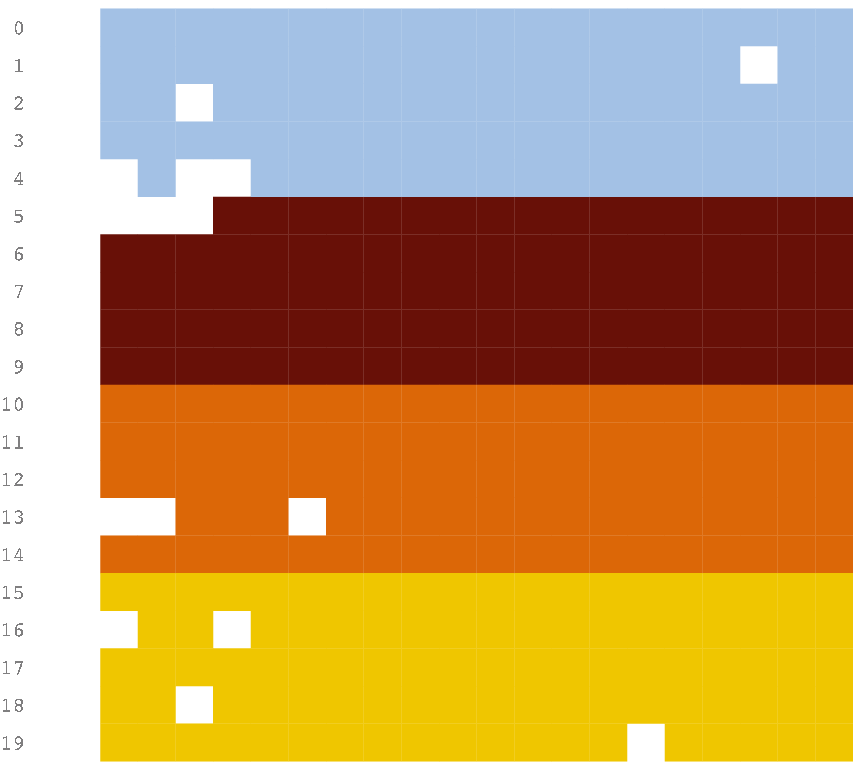}\hspace{.02\linewidth}}
\subfigure[$p=0.85$]{\includegraphics[width=.31\linewidth]{\figs/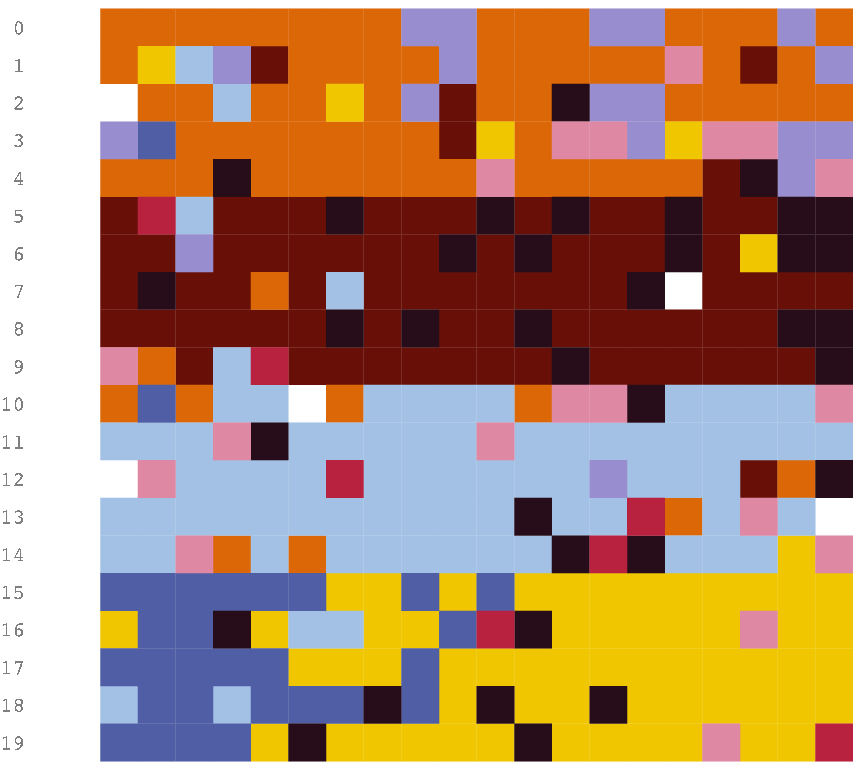}\hspace{.02\linewidth}}
\subfigure[$p=0.5$]{\includegraphics[width=.31\linewidth]{\figs/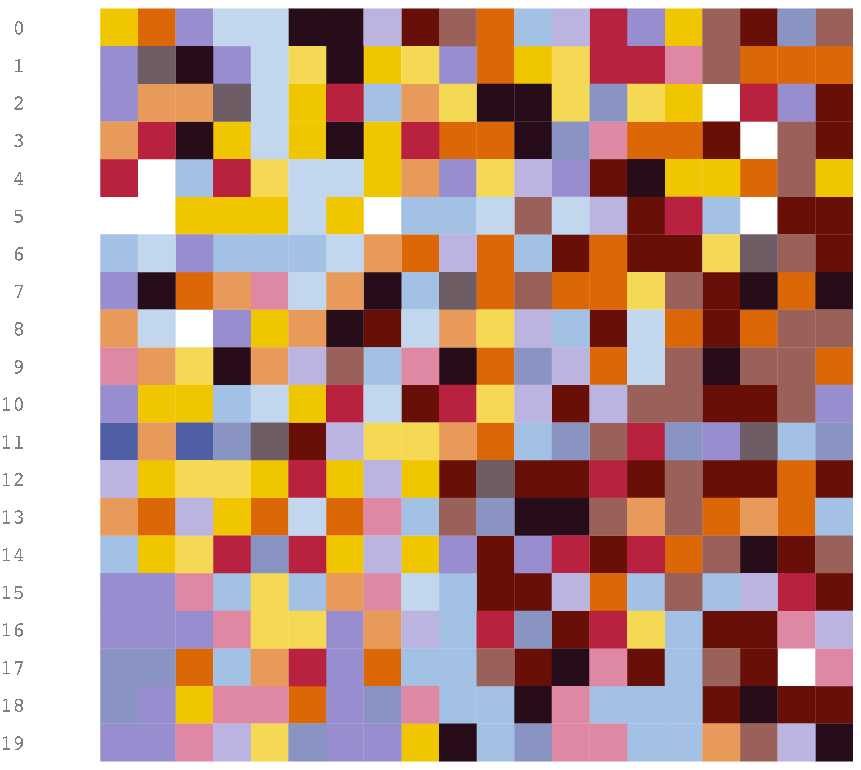}}
\vspace{-.5em}
\caption{Temporal graph $\GT$ constructed from the synthetic data, with \TB{$n_c=4$, $m=5$, \BM{$w=10$}, and $d=3$} for various probabilities of intra-community linking $p$. \TB{Left to right: $p=1$, $p=0.85$ and $p=0.5$, with respectively $5$, $9$ and $16$ detected temporal communities, each labeled using distinct colors; time goes from left to right, while physical nodes are spread on the y-axis (\#0 to \#19).}}
\label{fig:synthetic_data}
\end{figure*}

\begin{figure*}[h!]
\centering
\subfigure[Girvan-Newman]{\includegraphics[width=.31\linewidth]{\figs/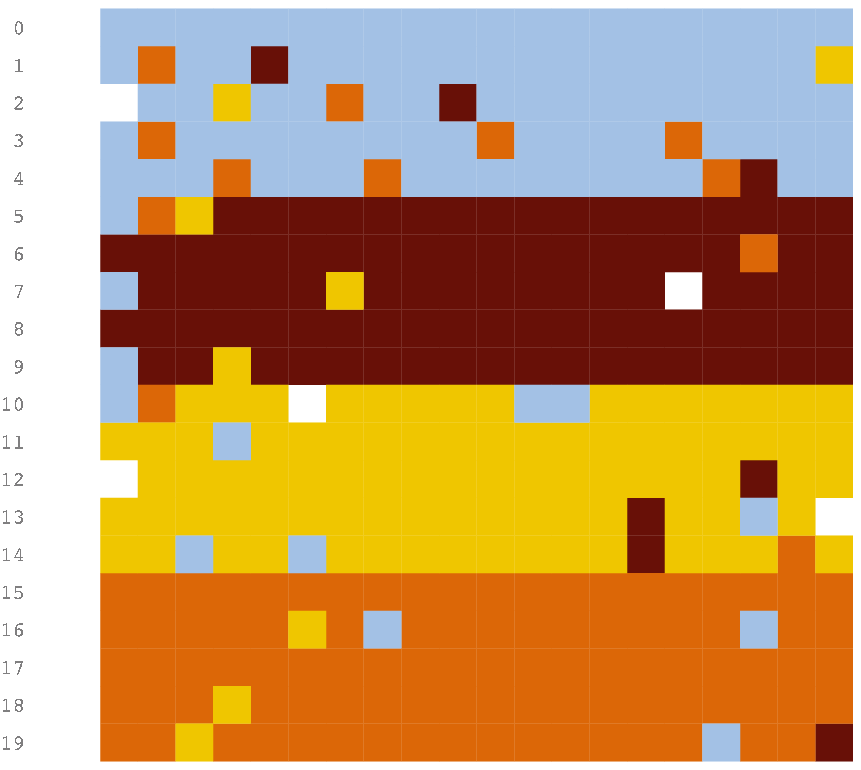}\hspace{.02\linewidth}}
\subfigure[Louvain {(same as above)}]{\includegraphics[width=.31\linewidth]{\figs/fig3-louvain-comm-p085.eps}\hspace{.02\linewidth}}
\subfigure[Walktrap]{\includegraphics[width=.31\linewidth]{\figs/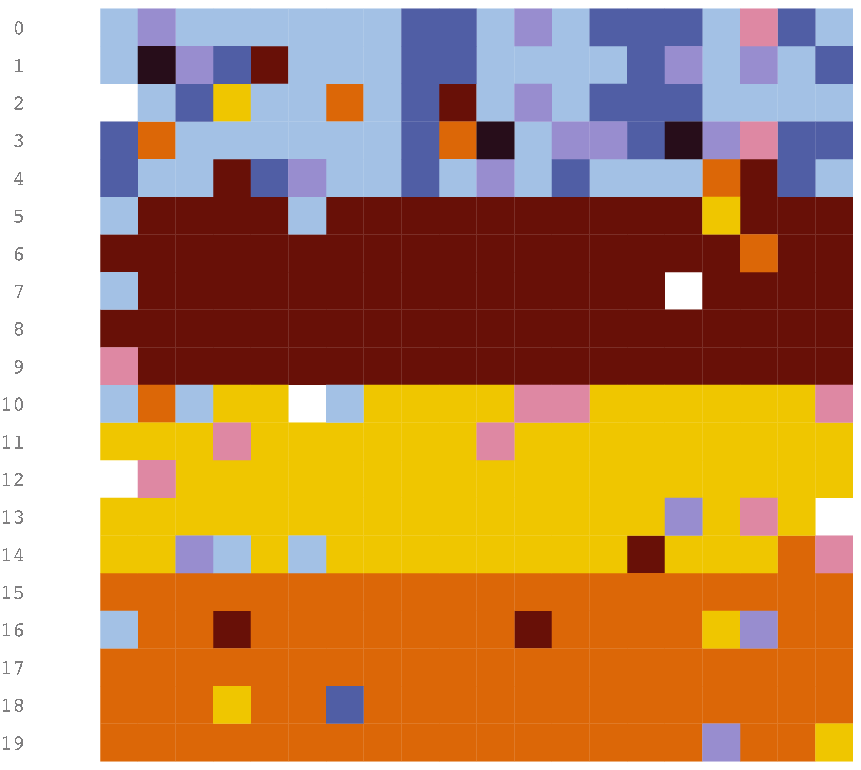}}
\vspace{-.5em}
\caption{Temporal communities detected for a unique synthetic dataset ($p=0.85$) using Louvain, Girvan-Newman and Walktrap algorithms.}
\label{fig:variousalgo}
\end{figure*}

Decreasing values of $p$ and $d$ lead to a temporal graph with decaying temporal community structure and recurrent interactions being more and more fortuitous. A totally random inter-community linking pattern thus corresponds to $p_{\text{random}}=\frac{1}{n_c}$.
To concretely illustrate this phenomenon and the functioning of our approach, we generate a synthetic citation data set $\Psi$ following the above-described procedure, with parameters: $n_c=4$ and $m=5$, $\tmax=20$, \BM{$w=10$} and $d=3$. Applying the Louvain community detection algorithm on the corresponding temporal graph yields temporal communities exhibiting a decaying temporal cohesiveness as much as $p$ decreases.
Figure~\ref{fig:synthetic_data} provides a simple illustration of how accurately the algorithm describes the temporal community structure, and especially matches the \emph{a priori} temporal communities when $p$ is high, \hbox{i.e.} when the data effectively obeys such a community structure.
In order to show the robustness of our methodology with respect to the chosen community detection algorithms, we present results with three distinct algorithms namely Louvain, Girvan-Newman and Walktrap, for the synthetic dataset, see Fig.~\ref{fig:variousalgo}. For the empirical dataset and for the remainder of this paper, we carry on with Louvain for its efficiency.

\subsection{Empirical data}

We now describe the two datasets we used to build empirical temporal graphs, namely scientific and blog citation data, respectively called {\bf ACL} and {\bf Blogs}.
\begin{itemize}
\item {\bf ACL.}

This dataset comes from the Association of Computational Linguistics (ACL).
It features the ACL Anthology Network -- a scientific bibliographical repository focused on computational linguistics\footnote{It may be browsed from \texttt{http://aclweb.org/anthology-new} -- see also \cite{Radev:2009:AAN:1699750.1699759}.}. The dataset is current as of Nov 2008.
It is paper-centered: it describes $13,706$ different papers, back to year 1965, such that we know that a paper published at some date cites some other papers published at some possibly earlier date.

We transform this data into a dynamic author-focused dataset featuring which author cites whom at which date, i.e. author $A$ cited at $t_A$ a production of author $B$ (possibly $A$ himself) at $t_B\leq t_A$.
The database contains a total of $247,059$ dynamic links among the $11,163$ authors of this 43-year collection.

\item {\bf Blogs.}

This dataset comes from a collection from February 1$^{st}$ to July 1$^{st}$, 2010, of blog posts of a significant part of the active French-speaking blogosphere, dynamically monitored by {\sc Linkfluence}, a company specialized in digital community surveys.

It features post contents, most importantly {\em who publishes a post, when, which other post this post does cite, and when that other post was published}.
In other words, it features links of the form blog $A$ cites at $t_A$ a post published by blog $B$ (possibly the same as $A$) at time $t_B\leq t_A$.
We get 49,232 dynamic links among 10,436 blogs during the 151-days crawling period.

\end{itemize}

\section{Results}\label{sec:results}

\subsection{Qualitative insights}


\paragraph{Topical clustering} As expected, communities found on both datasets exhibit a certain consistency regarding their topic: for instance, one large community in ACL principally includes authors of papers dealing with ``statistical machine translation''. As such, it would surely be comforting that temporal communities are consistent with the type of results that static algorithms would usually yield: let us therefore focus on situations specific to temporal communities 
which are likely to be less accessible through classical static community detection approaches.

\paragraph{Events} Communities related to specific events, and therefore characterized by a relative burst of interactions, are also exhibited. For instance, the Blogs dataset features 16 temporal communities which all involve the same core of 3 specific nodes, and sometimes an extra other node. The associated timestamps for these communities are quite regular: each community is separated of the next one by around one week. A closer examination of the database informs us that these communities correspond to the weekly ``best of'' made by blog \url{filmgeek.fr} --- specialized in movie reviews --- which often involves the same group of blogs: \url{blogywoodland.blogspot.com}, \url{cinefeed.com} and, more rarely, \url{toujoursraison.com}.
Such quasi-periodic phenomenon is a clear evidence of a specific, recurrent event.

\paragraph{Strong recurrent interactions} The temporal approach makes it possible to easily detect interacting ``cores'' without introducing extra hypotheses on link weights --- ``cores'' made of, for instance, a small group of strongly interacting nodes, with or without a ``periphery'' of other weakly interacting nodes. 
For example, ACL features some small temporal communities consisting of several authors sharing a particularly strong bond, such as a group of 3 Japanese authors (namely S. Yoshida, T. Hitaka and H. Tsurumaru) who published several times together, also citing their previous papers.
With a static, unweighted approach, these cores may be overlooked as usual elements of a possibly larger community.




\subsection{Temporal metrics and dynamic community profiles}\label{sec:metric}\label{sec:temporalmetrics}

We now introduce various temporal metrics in order,
and thereby distinguish temporal communities. We show how these metrics can be used to focus on specific types of temporal communities and, at a higher level, how they can be used to discriminate  empirical contexts through distinct dynamic community profiles.

We essentially suggest that the following dimensions are key in appraising temporal community types:
(1) Community size (\hbox{i.e.} extent of the community within the system); (2) Recurrence of interactions; (3) Interactivity (and self-citation); and (4) Balance (\hbox{i.e.} connectedness heterogeneity). 



\paragraph{Community size ($z$)}
Temporal community extent may be measured through the {community size}, which we define in terms of physical nodes. More formally, $z(C)$ denotes the number of unique nodes participating in the temporal community: \begin{equation}\forall C\in\CT,\quad z(C)=\Big|\{v_i\,|\,\exists (v_i,t_i)\in C\}\Big|\end{equation}On Fig.~\ref{fig:NA}, communities ACL \#1 and \#26 are respectively of size $3$ and $9$.

\paragraph{Recurrence of interactions: node activity $NA$}

\newcommand{\LD}{\text{LD}}
\newcommand{\NA}{\text{NA}}
\newcommand{\LR}{\text{LR}}
\newcommand{\SC}{\text{SC}}
\newcommand{\HI}{\text{HI}}
We characterize the recurrence of node participation in a temporal community in terms of node activity $\NA$.
It is a normalized index simply based on the ratio of temporal nodes with respect to physical nodes in the community. Formally, we define: \begin{equation}\label{eq:td}\forall C\in\CT, \quad\NA(C)=1-\frac{z(C)}{|C|}\end{equation}
Hence, a highly active community is more likely to feature a small number of physical nodes (low $z(C)$) appearing repeatedly over several timestamps (high $|C|$), resulting in a $\NA$ close to 1.
For instance, on Fig.~\ref{fig:NA} we see that temporal community ACL \#1 has a high $\NA$, with only three nodes quite recurrently interacting with each other.\footnote{We may also represent node recurrence within a given temporal community by (i) using the average degree in the corresponding temporal subgraph, \hbox{i.e.} using the ratio between $|C|$ and the number of temporal links, or (ii) by using repeated appearances of physical links, \hbox{i.e.} the ratio between (the number of temporal links of the temporal subgraph) and (the number of links in the projection of corresponding temporal subgraph projection onto a static, unweighted network).
However, we find that the former is empirically correlated to community size $z(C)$ while the latter is fairly correlated to $\NA(C)$.}
\TB{Note also that the Louvain algorithm captures cohesive temporal subgraphs, i.e. dense patterns of either recurrent or multiple interactions. 
As we see in Fig.~\ref{fig:NA}, low NA communities exhibit an increased branching structure and correspond more to star-like multiple interactions than recurrent ones.  In this respect, NA filters temporally recurrent patterns among those detected by the Louvain algorithm.}
\begin{figure}[h!]
\centering\small\begin{tabular}{ccc}
\multicolumn{3}{c}{\bf NA}\\
high&~\hspace{2em}~&low\\\midrule
\includegraphics[width=.425\linewidth]{\figs/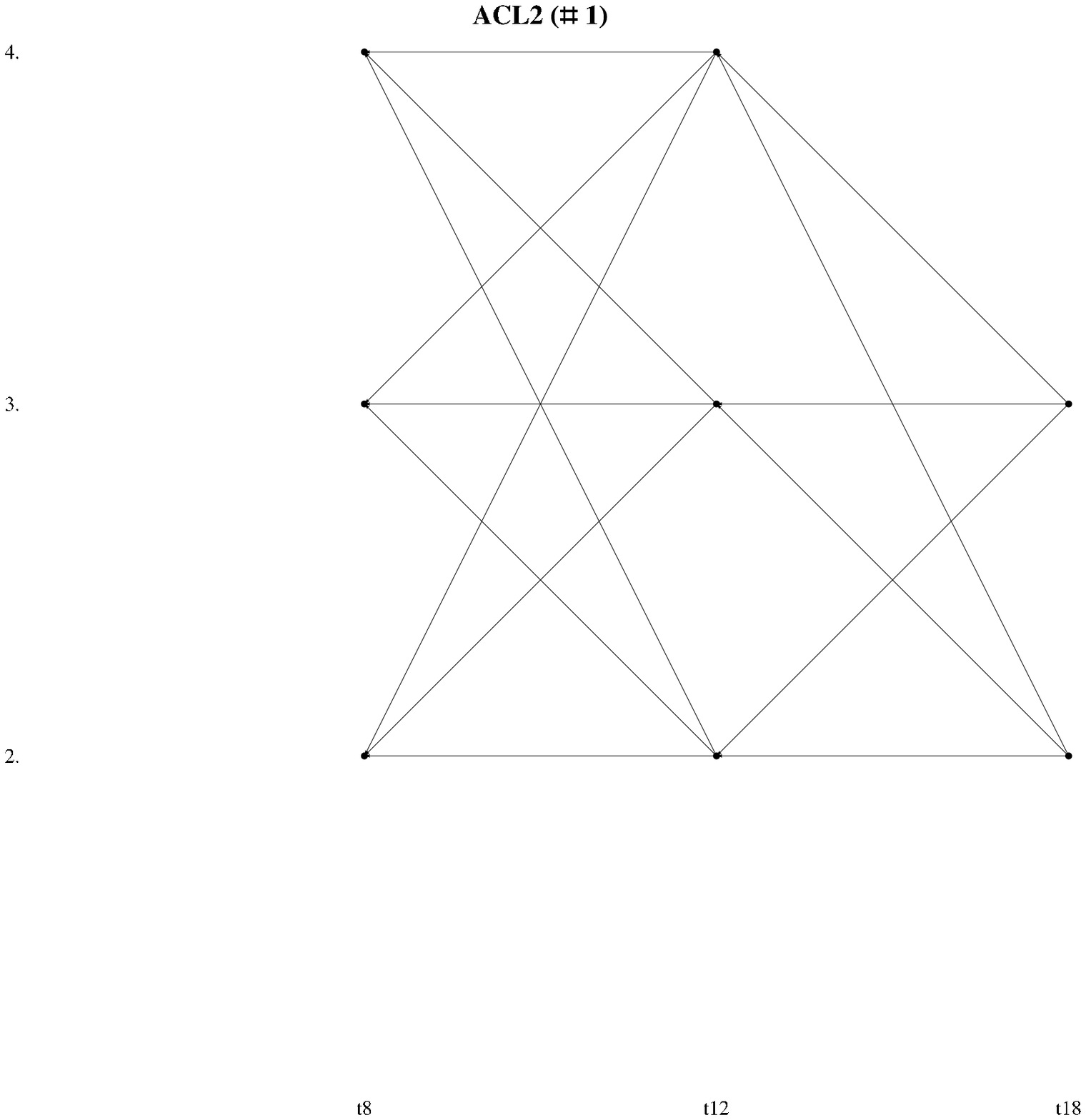}
&&\includegraphics[width=.22\linewidth]{\figs/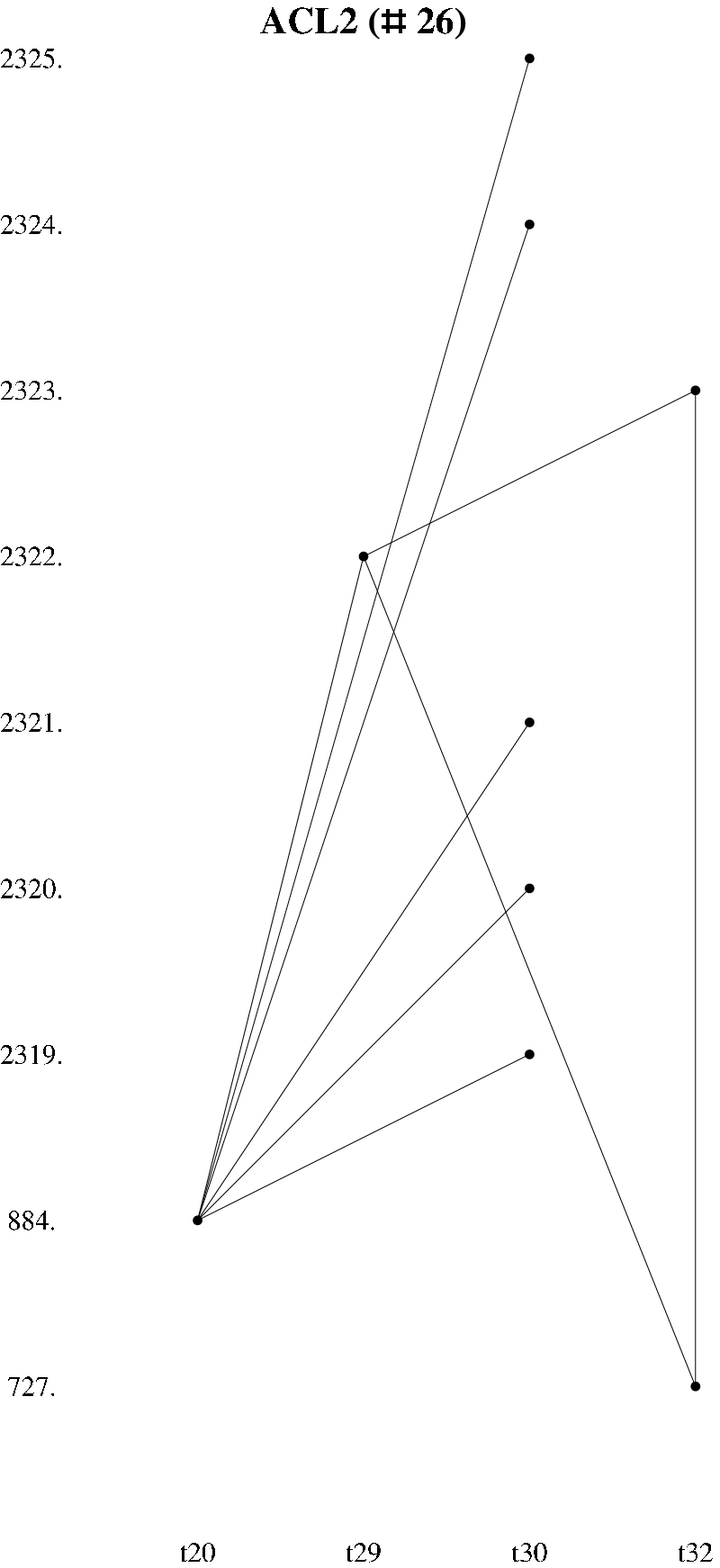}\hspace{1em}
\end{tabular}
\caption{\label{fig:NA}ACL data with two typical communities having respectively a high (\emph{left}) and low (\emph{right}) node activity \NA.}
\end{figure}

\begin{figure}[b!]
\centering\small
\begin{tabular}{m{.2in}>{\centering}m{.26in}|>{\centering}%
  m{.5\linewidth}>{\centering\arraybackslash}m{.3\linewidth}}
&\multicolumn{1}{c}{}&\multicolumn{2}{c}{\bf SC}\\
&\multicolumn{1}{c}{}&high&low\\\cmidrule{3-4}
\multirow{8}{*}{\bf HI}&high&
\includegraphics[width=\linewidth]{\figs/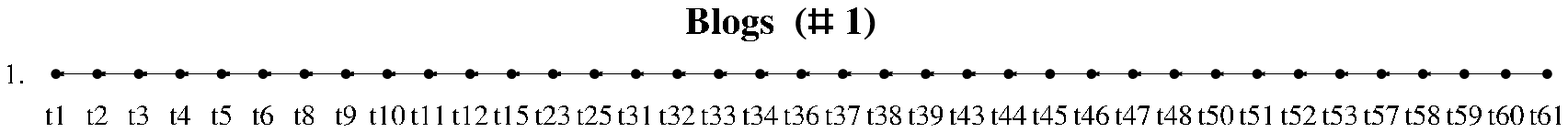}&
\includegraphics[width=.7\linewidth]{\figs/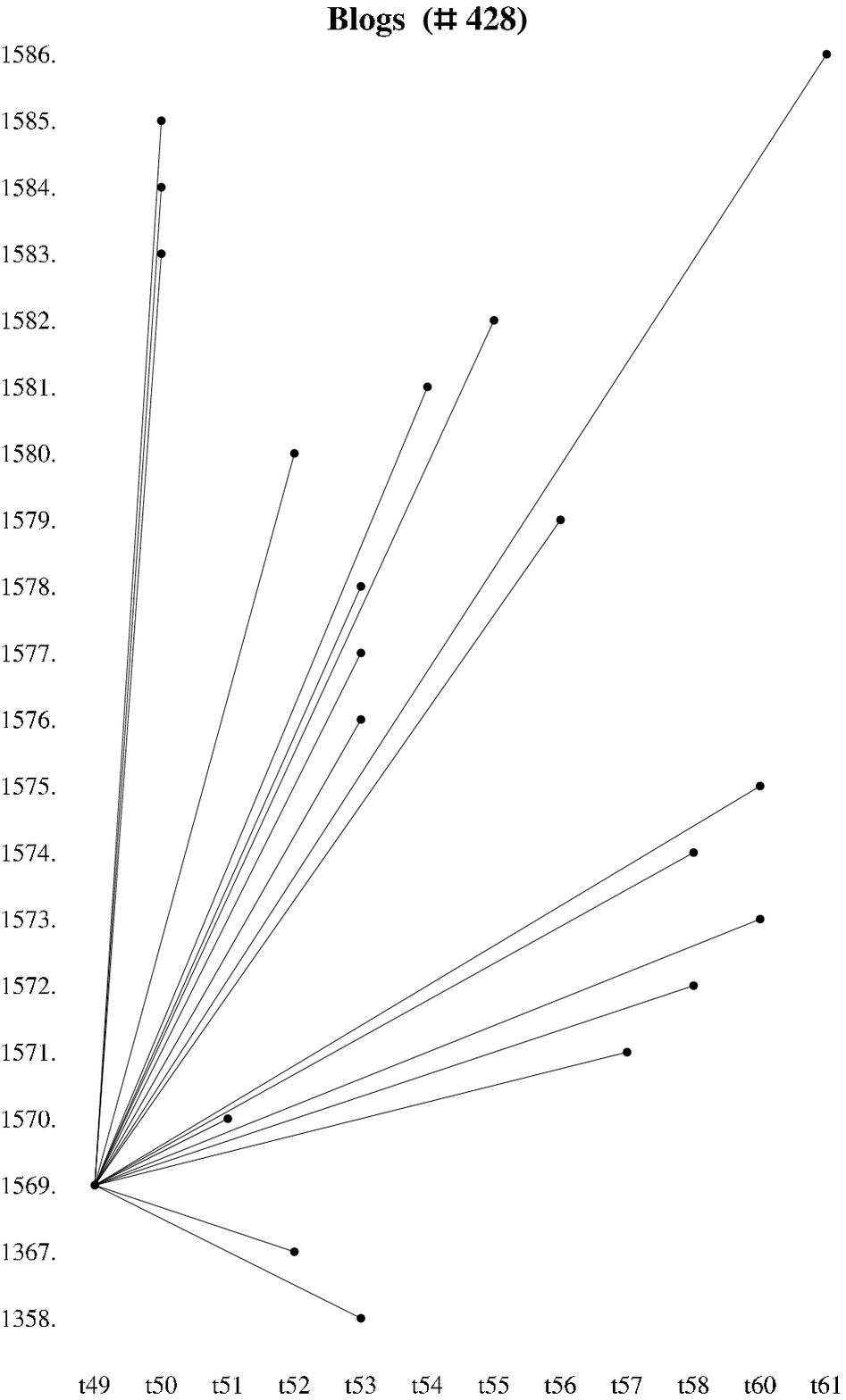}\\
&&\\
&low&
\includegraphics[width=\linewidth]{\figs/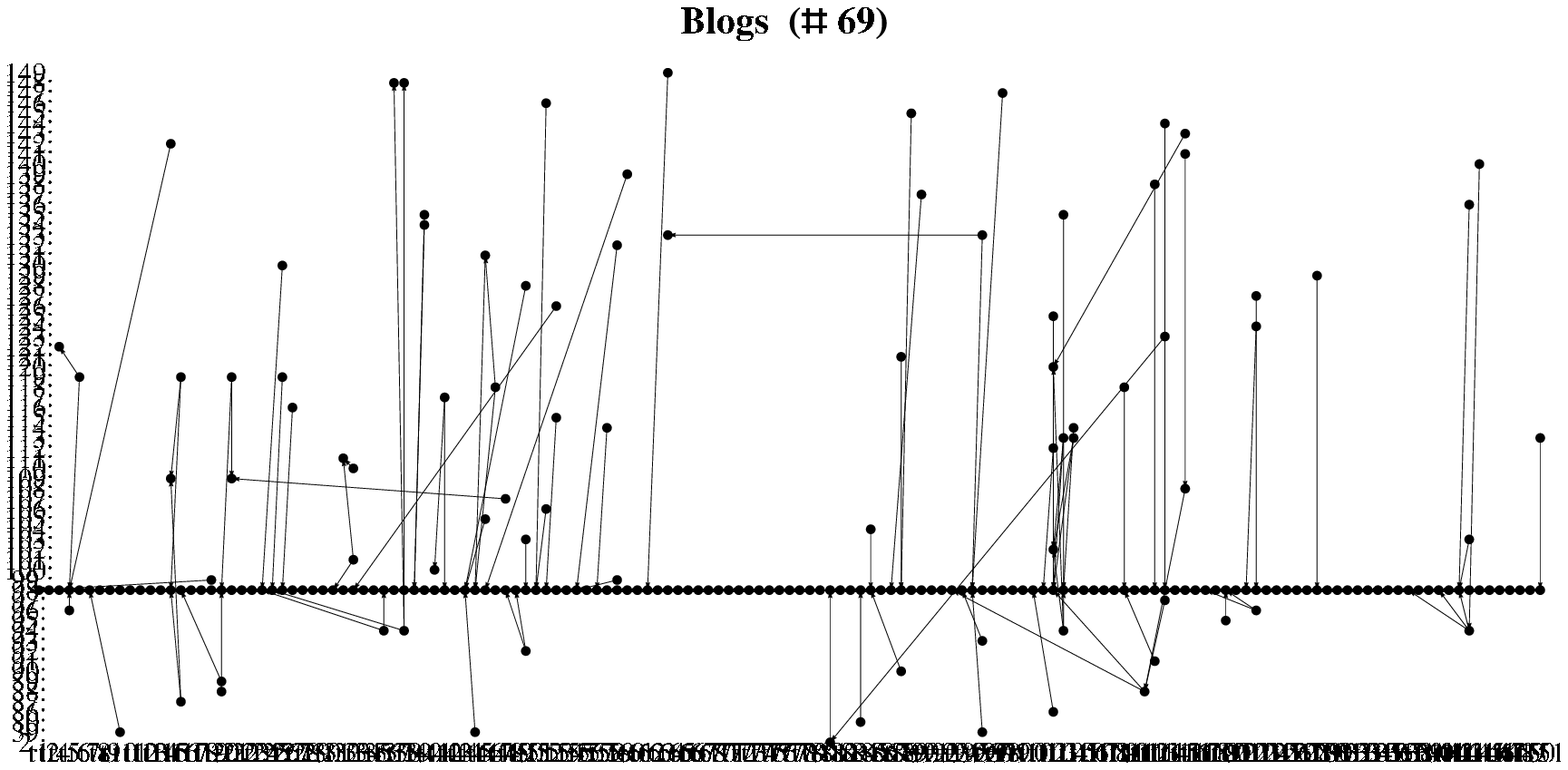}&
\includegraphics[width=.7\linewidth]{\figs/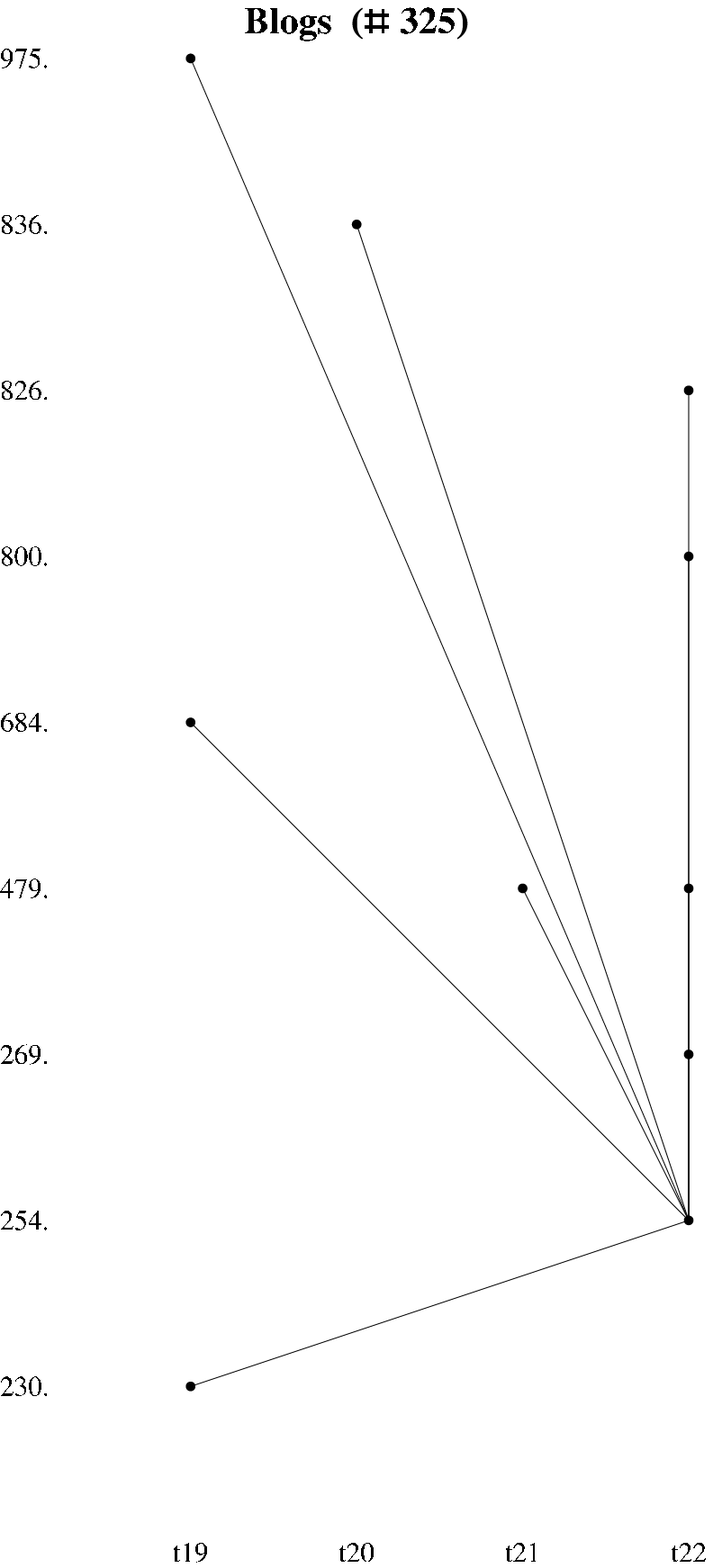}
\end{tabular}
\caption{Typical temporal community patterns with respect to $\SC$ and $\HI$ for ``Blogs''.  When \SC{} is high, temporal communities could look like a long self-citation line (high \HI, \#1) or a backbone (low HI\, \#69); when \SC{} is low, they could appear as incoming stars (high \HI, \#428) or  outgoing stars (low \HI, \#325).}\label{fig:temoral_panel}
\end{figure}

\paragraph{Interactivity: self-citation ratio}
A temporal community may contain a large number of temporal nodes and links while essentially consisting of a node citing repeatedly itself (a relatively frequent pattern on the blogosphere and, to a lesser extent, in science).  
The self-citation ratio $\SC(C)$ is defined as the ratio between the number of self-citations and the total number of temporal links present in the community. Formally,
\begin{equation}\forall C\in\CT,\quad\SC(C)=\frac{|\{((v_i,t_i),(v_i,t_j))\in C^2\}|}{|\{((v_i,t_i),(v_j,t_j))\in C^2\}|}\end{equation}
This measure primarily targets temporal communities where a pivotal node intensely cites itself over a large number of timestamps, possibly being sparsely cited by other surrounding nodes --- compare 
high-$\SC$ communities ``Blogs~\#1'' with ``Blogs~\#69'' on Figure~\ref{fig:temoral_panel}.

\paragraph{Balance: heterogeneity of interactions}Temporal communities, as can be seen on Fig.~\ref{fig:temoral_panel}, may possess balanced or unbalanced activity, depending on whether one physical node much more heavily and repeatedly cites other, or all physically nodes eventually cite each other in an even manner. 
We propose to measure this structural heterogeneity using an index based on Herfindahl's index $h$:
\begin{equation}h(C)=\frac{1}{z(C)\sum_{i}^{z(C)}{p_i^2}}\end{equation} where $p_i(C)$ is the normalized number of temporal links stemming from a given physical node $i$ within $C$ (it is normalized such that $\sum_i^{z(C)} p_i(C)=1$, $z(C)$ being the size of community $C$). Because $h(C)\in[1/z(C),1]$, we eventually rescale it into $[0,1]$: \begin{equation}\HI(C)=\frac{z(C)h(C)-1}{z(C)-1}\end{equation}
Put shortly, \HI{} is closer to 1 when temporal communities have an homogeneous out-going connectivity (e.g. ``Blogs \#1'', as opposed to ``Blogs \#325'').

%
%

\subsection{Dynamic profiles}

\begin{figure}[h]
\centering
\subfigure[Temporal metrics $\NA$, $\SC$, and $\HI$ for synthetic data, with respect to intracommunity probability $p$. (Inset: average size of communities in terms of physical nodes.) \label{fig:synth_q}\label{fig:tempo_q}]{\includegraphics[width=.45\linewidth]{\figs/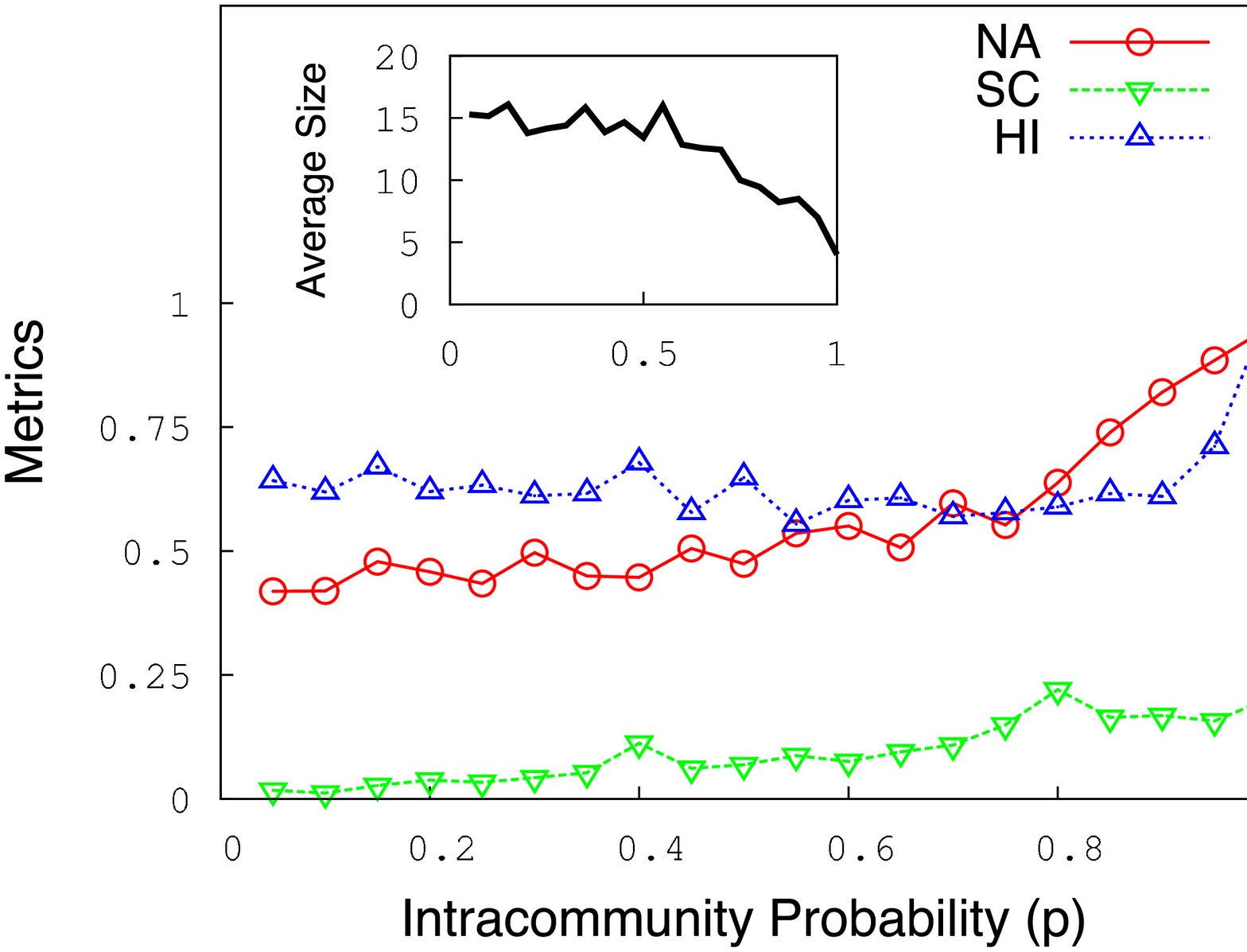}}
\hspace{.02\linewidth}
\subfigure[\label{fig:quality}Dissimilarity$^\text{\ref{fn:1}}$ between \emph{a priori} community assignment and \emph{ex post} detected communities. 
]{\includegraphics[width=.45\linewidth]{\figs/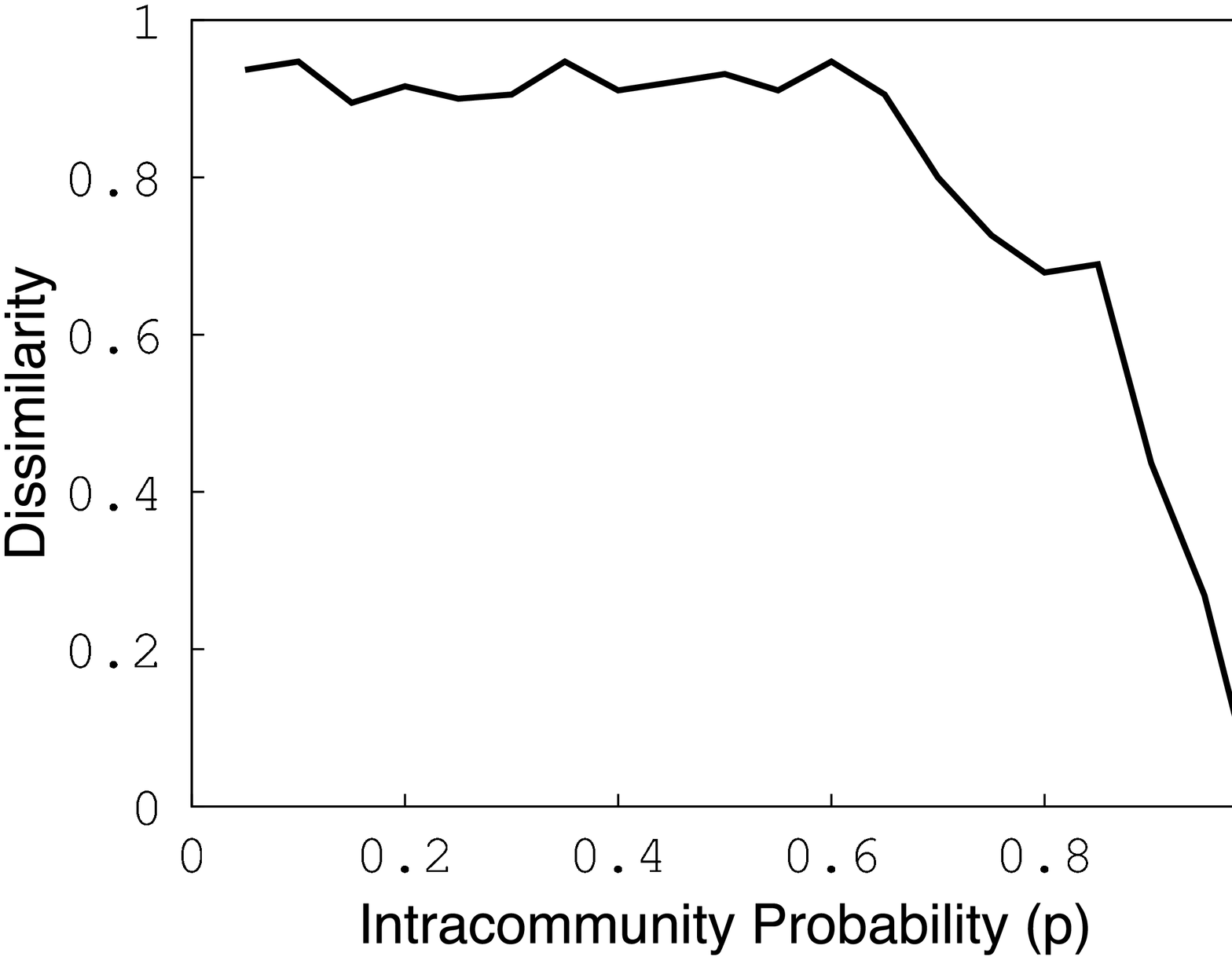}}
\caption{Behavior of metrics on the synthetic toy model \TB{with respect to intracommunity probability $p$}.}
\end{figure}

\paragraph{First glimpse: synthetic toy model} Fig.~\ref{fig:tempo_q} shows the behavior of the aforesaid metrics in synthetic data for various values of the inter-community linking probability $p$. As can be observed, these metrics are unaffected for most values of $p$, except for the highest values where all metrics $\NA$, $\HI$ and $\SC$ increase relatively sharply. Inset of Fig.~\ref{fig:tempo_q} shows that in the weakly cohesive network, temporal communities have an average size which is close to the number of nodes, \hbox{i.e.} they all  tend to be  spread over a large portion of the physical node set. By contrast, temporal communities for larger values of $p$ are much smaller, and indeed closer to the \emph{a priori} community attribution, \hbox{i.e.} they are much more focused on a small subset of nodes.

This behavior echoes the match between \emph{a priori} community assignations and actual node interaction profiles for a given $p$. Obviously, the original community assignation is likely to vanish as $p$ diverges from $1$. Yet, it actually vanishes at around a range of values of $p$ which is similar to the region where the behavior of the above metrics change, as shown on Fig.~\ref{fig:tempo_q}. Indeed, if we define a dissimilarity quantity expressing how much the \emph{ex post} community detection diverges from the original a priori community assignation\footnote{\label{fn:1}More precisely: given $(C^-)_{i=1,..,mn_c\tmax}$ a vector such that $C^-_i$ represents the \emph{a priori} community of temporal node $i$, and  $C^+$ a same-length vector of detected communities, we consider $\label{dis} d=\sum_{i,j}\left(\delta(C^-_i,C^-_j) \ \mathsf{XOR} \ \delta(C^+_i,C^+_j)\right)$, where $\delta(x,y)$ is the Kronecker function. Put simply, if a pair of temporal nodes is classified differently by the \emph{a priori} assignment and the \emph{ex post} detection result, $d$ will be increased by 1.
We eventually consider a dissimilarity quantity $D$ which is $d$ normalized with respect to all possible node combinations in the network.}, we indeed notice that there is a sharp change in dissimilarity  as $p$ reaches the region around $[0.8,0.9]$. See Fig.~\ref{fig:quality} for an illustration.

\begin{figure}[!t]
\centering
\begin{tabular}{cc}\toprule
\multicolumn{2}{c}{\rm \scriptsize Synthetic}\\
\includegraphics[width=.47\linewidth]{\figs/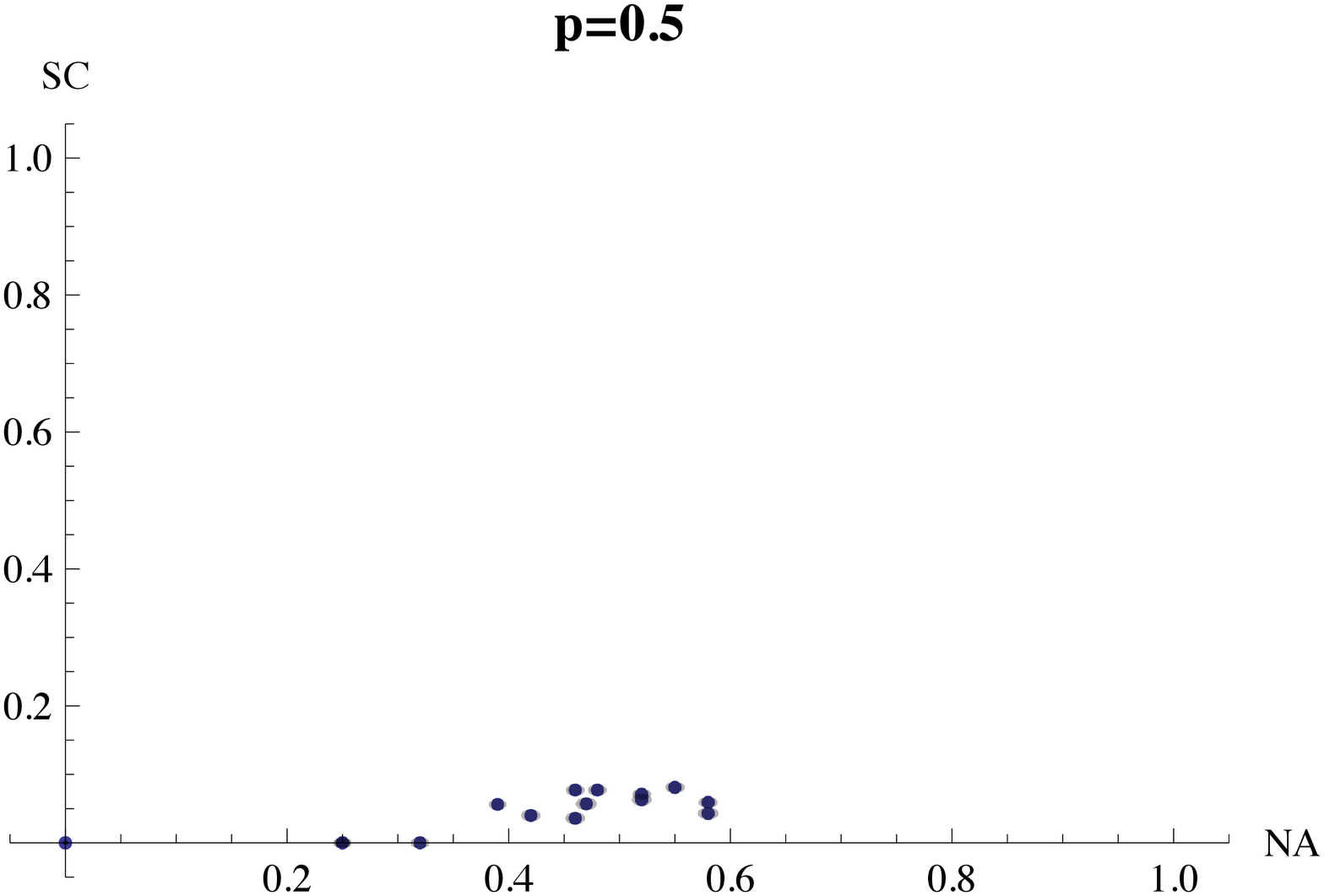}&
\includegraphics[width=.47\linewidth]{\figs/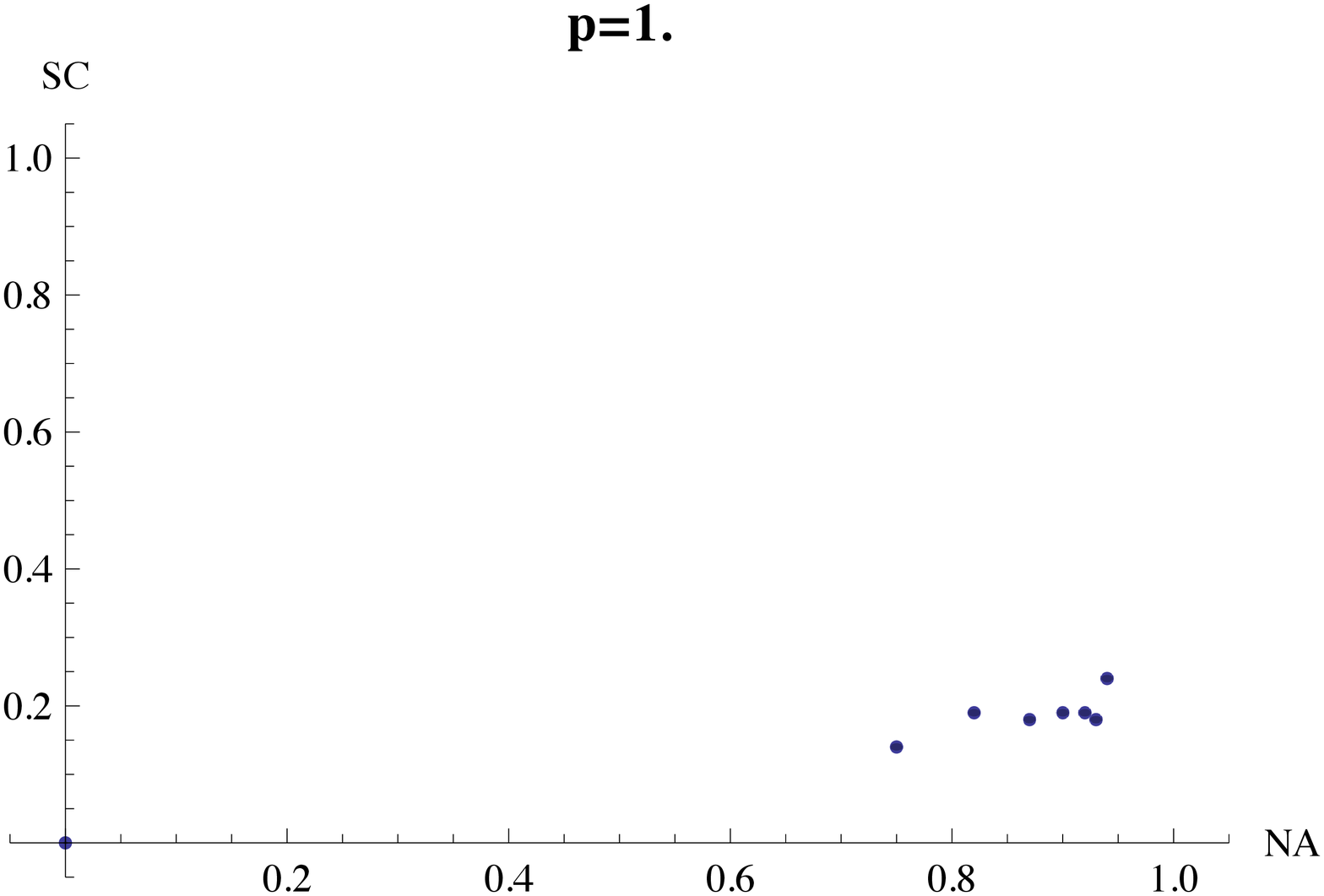}\\
\midrule
\multicolumn{2}{c}{\rm \scriptsize Empirical}\\
\includegraphics[width=.47\linewidth]{\figs/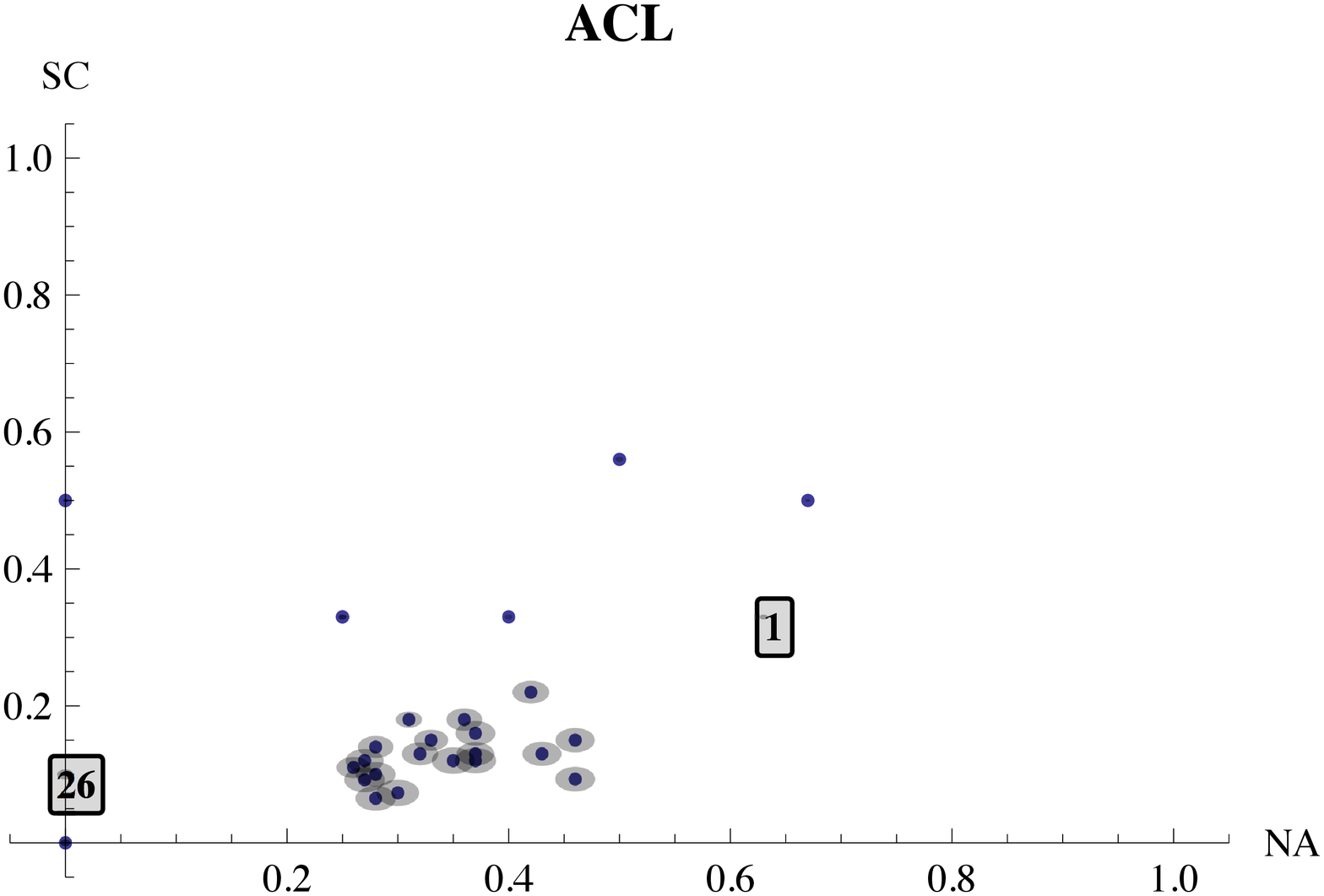}&
\includegraphics[width=.47\linewidth]{\figs/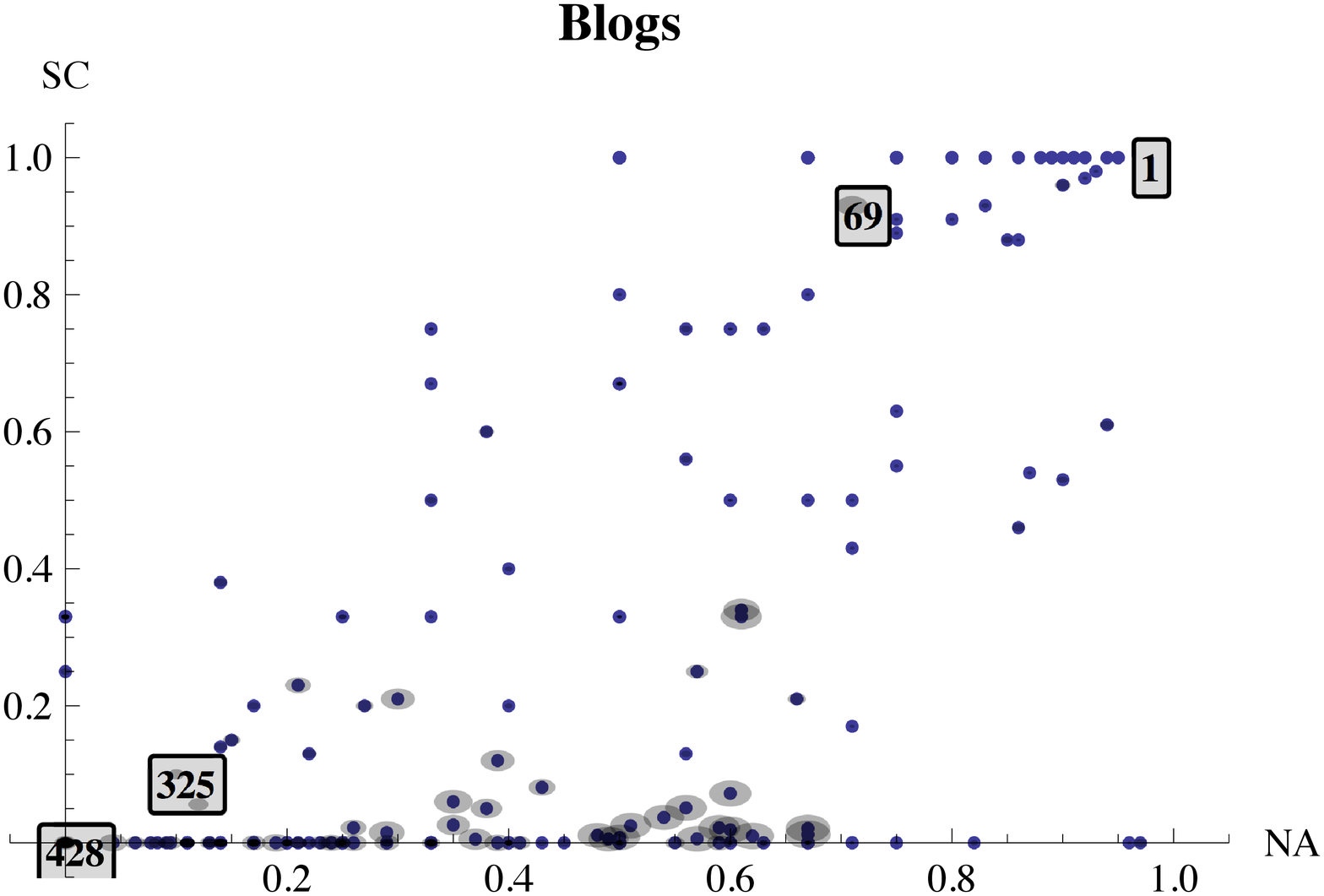}\\
\bottomrule
\end{tabular}
\caption{\label{fig:profiles}Position of temporal communities for, on \emph{top}, synthetic data ($p=0.5$ and $p=1$) and, at \emph{bottom}, ACL and Blogs.  Global profiles are plotted according to $\NA$ and $\SC$ values (resp. x-axis and y-axis) and their size (represented by a gray disk whose area is logarithmically proportional to $z(C)$). Framed labels correspond to temporal communities represented in the previous figures.
}
\end{figure}

\paragraph{Global temporal landscape} We may now combine some of these temporal metrics to depict general dynamic profiles for each dataset.  Fig.~\ref{fig:profiles} represents the positions of all temporal communities in a $\SC$ \hbox{vs.} $\NA$ plane, together with their size.  (Other dimensions could be selected, depending on the focus.) Strong discrepancies appear for each datasets. The synthetic dataset, for instance, exhibits a progression from low- to high- $\NA$ temporal community population as $p$ goes to 1, confirming the a priori knowledge that datasets built with higher $p$ values are indeed yielding more cohesive temporal communities. Self-citation is relatively rare however in that context, compared with the empirical dataset profiles, especially blogs.

In this respect, ACL is principally made of larger temporal communities with globally medium $\NA$ indices and non-negligible $\SC$ indices (indicating that most of these scientific citation temporal communities are made of a bit of self-citation embedded into a wider web of recurrent non-selfcitations), and a small number of either zero-$\NA$ communities (which are basically negligible communities corresponding to one instance of a citation of B by A) or higher $\SC$ and $\NA$ communities (which are instances of self-citation of a group of authors across a small number of timesteps).  On the other hand, Blogs display a much more contrasted profile: there is a large number of highly self-citing communities, with moderate to high $\NA$, and a large number of non-self-interacting communities, with low to medium-high $\NA$.

Put differently, Fig.~\ref{fig:profiles} provides a global overview of the temporal interaction profile of each dataset, plausibly typical of the underlying referencing behavior. It also emphasizes that a single all-purpose temporal community definition is likely to overlook the diversity of recurrent interaction contexts, as datasets are structurally significantly different from one another and rely upon distinct inter-temporal interaction behaviors.

\subsection{Impact of graph density / sparsity}
\begin{figure*}
\centering
\subfigure[$d=2$]{\includegraphics[width=.33\linewidth]{\figs/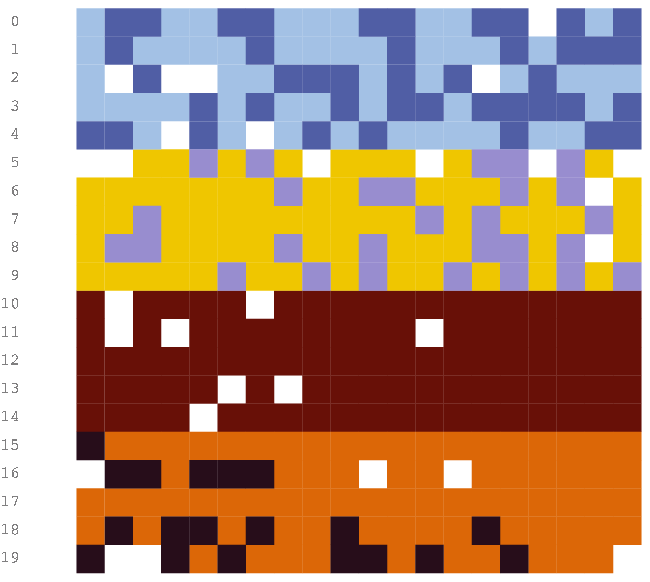}}\hspace{1cm}
\subfigure[$d=4$]{\includegraphics[width=.33\linewidth]{\figs/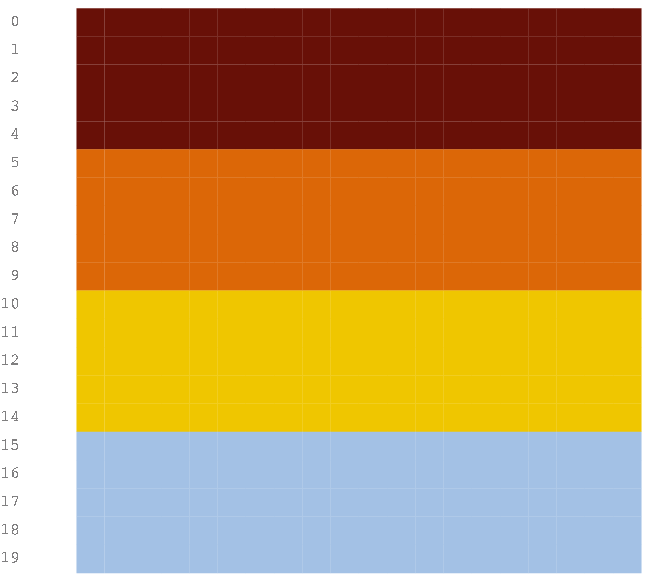}}
\vspace{-0.1in}\caption{\label{fig:deg_2}\label{fig:deg_4}\label{fig:density}Visual representation of the dynamic community structure \TB{of synthetic datasets created using various average degrees $d$, and a maximal intracommunity probability $p=1$}.}
\end{figure*}

The level of interaction intensity is likely to affect temporal community structure: admittedly, a sparse graph will exhibit smaller communities, in a more fragmented way. For instance, coming back to the toy synthetic data, a decrease in the average degree/density leads to a less coherent temporal community landscape:
for example, in Fig.~\ref{fig:density}, we illustrate that in a synthetic dataset with identical parameters as before, except for  an average degree of $2$, leads to as many as $17$ communities even for $p=1$ (Fig.~\ref{fig:deg_2}-left), while an average degree of $4$ leads assuredly to the $4$ a priori communities (Fig.~\ref{fig:deg_4}-right).

This issue essentially boils down to the fact that sparse graphs may lead to sets of small meaningless communities. It is an issue which is certainly common to community detection methods, and it is typically addressed by an increase in the granularity of the input data, for instance by aggregating data on longer time periods or considering a lower time sampling rate; or by eventually merging communities based on various criteria. Solving this issue in general is beyond the scope of this paper; yet, we may discuss several ways of addressing it in our specific context. In particular:
\begin{itemize}
\item The temporal grain may be increased in order to take into account a slower pace of linking.
This would not be equivalent to longitudinal approaches, as we would still rely on the inter-temporal intertwinement of links across various periods to build communities.
For instance, our results on ACL and blogs are robust when time sampling is halved, i.e. when considering that year 1980 and 1981 correspond to a single timestamp, as do 1982 and 1983, but not 1981 and 1982.
\item Another way of proceeding would be to merge some communities based on the temporal information, using temporal nodes of a single physical node which are present at close timesteps, in a fashion not unfamiliar to what has been done in \cite{jdidia,mucha} by adding inter-temporal links across temporal nodes. Similarly, temporal metrics of Sec.~\ref{sec:temporalmetrics} may be used as guidelines to filter out irrelevant temporal communities.
\TB{We will sketch a few preliminary details of this latter approach in Sec.~\ref{sec:merge}.}
\end{itemize}


\begin{figure*}
\begin{center}
\subfigure[Node lifetime distribution for Blogs (scale in days), inset: for ACL (scale in years).\label{fig:node_lifetime_dist}]{\includegraphics[width=3.8cm]{\figs/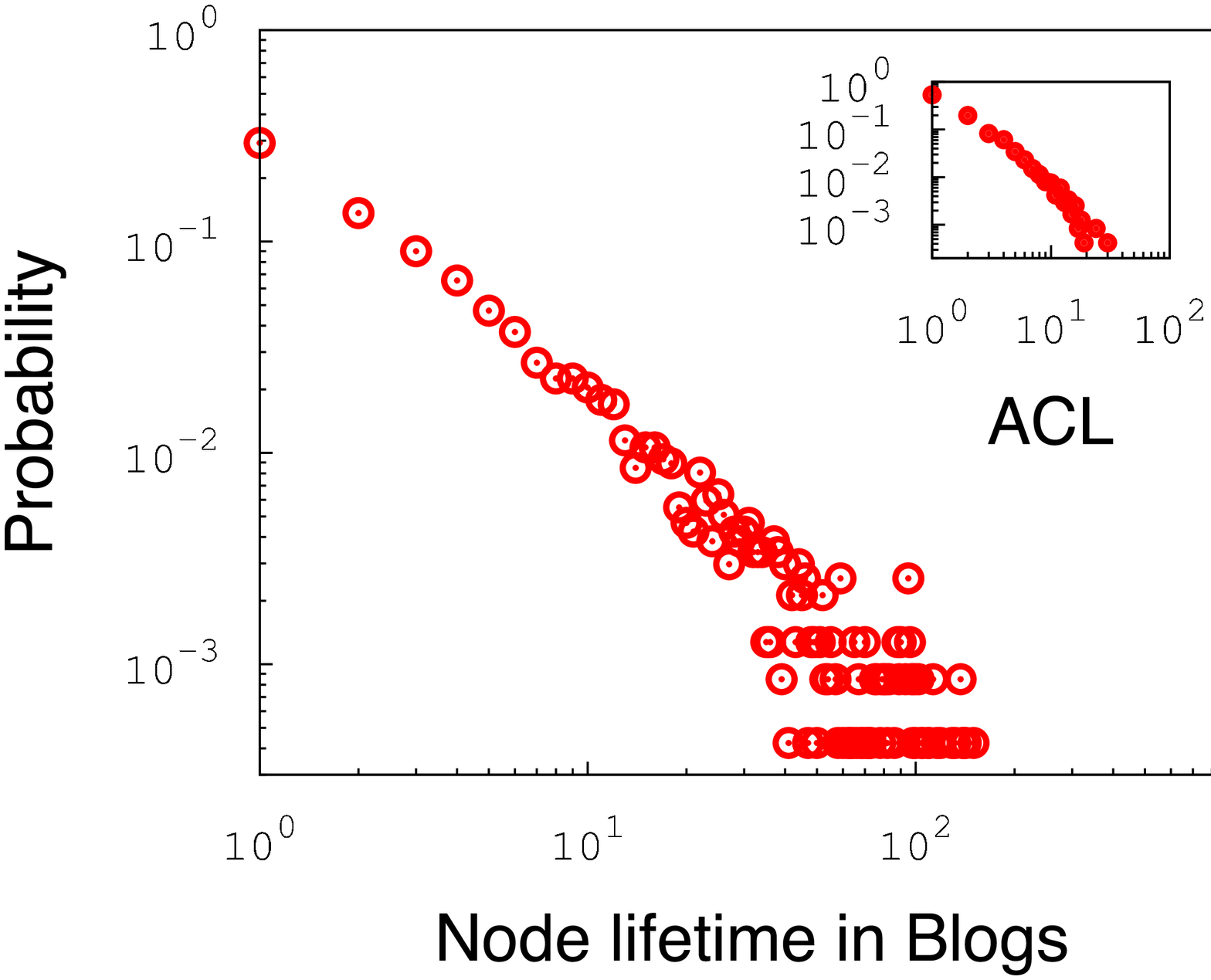}}\hspace{.02\linewidth}
\subfigure[Community membership distribution for Blogs (scale in days), inset: for ACL (scale in years).]{\label{fig:comm_mem_dist}\includegraphics[width=3.8cm]{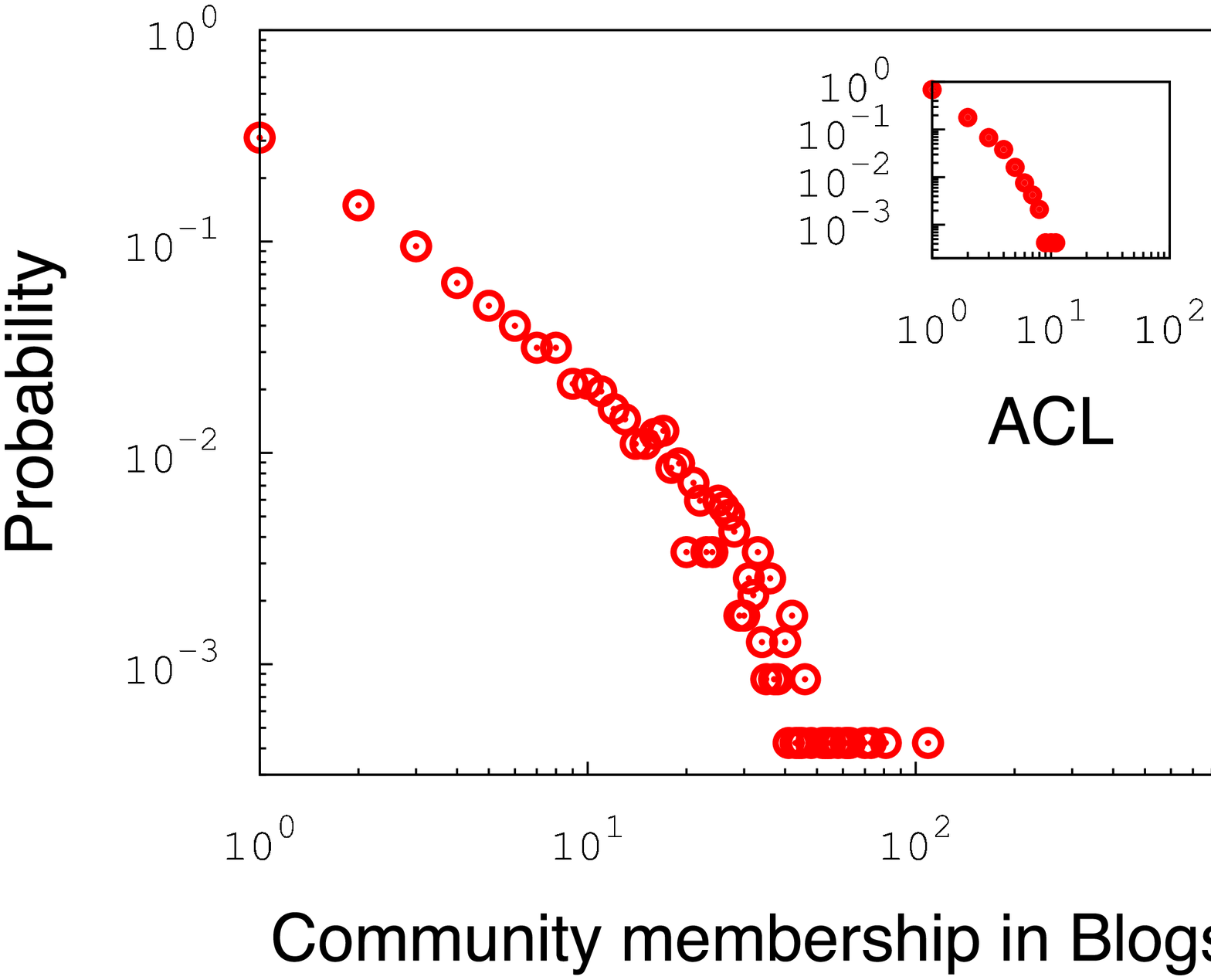}}\hspace{.02\linewidth}
\subfigure[Correlation between node lifetime and community membership for Blogs; inset: for ACL.]{\label{fig:n_comm_lifetime_assort}\includegraphics[width=4.8cm]{\figs/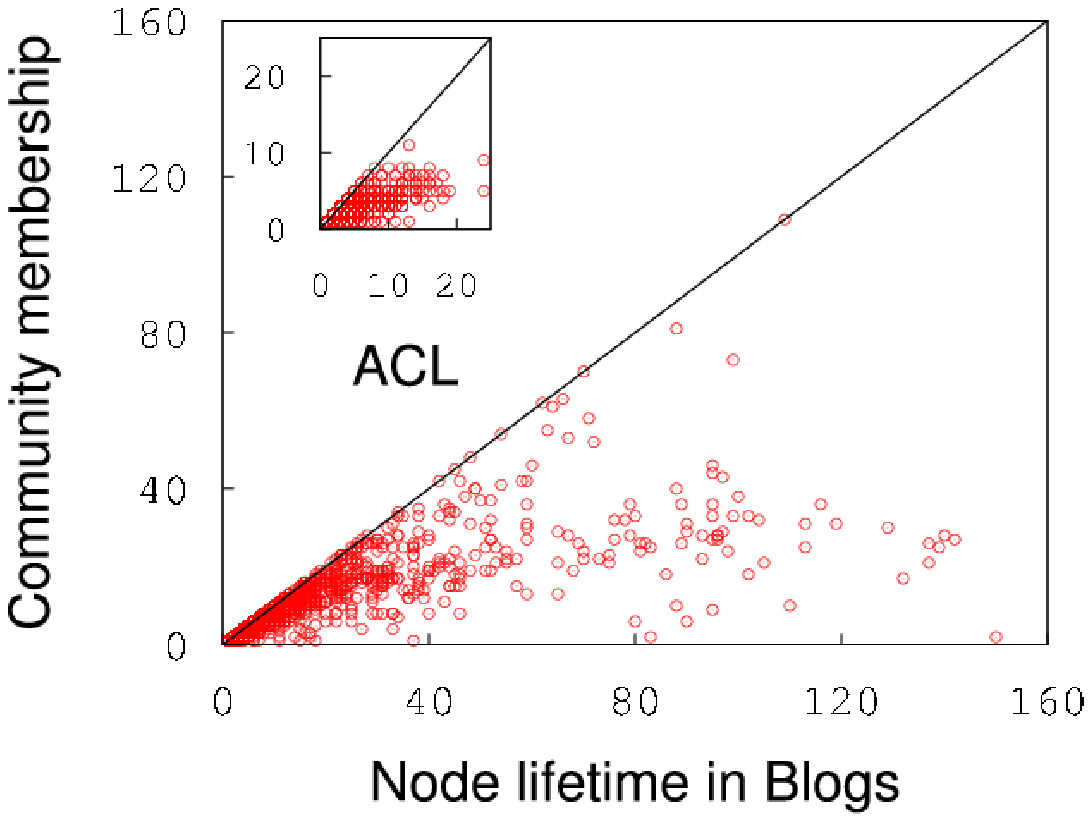}}
\end{center}
\vspace{-0.2in}
\caption{Distributions of node lifetimes (number of distinct timesteps of activity in the dataset), community memberships (number of temporal communities a given node belongs to, over the whole dataset), and related correlations.} \label{fig:mambership}
\end{figure*}

\section{Behavior of nodes inside and across temporal communities}\label{sec:behavior}
The description of temporal communities provides a landscape of the various foci of activity of physical nodes across time: some nodes may be devoted to a given subgroup of nodes, while others may frequently jump from a community to another. We now propose a few simple metrics to measure community membership parameters of individual nodes, and use them on the ``Blogs'' and ``ACL'' datasets to characterize different bloggers and authors, as well as the temporal cohesiveness of the respective underlying systems as a whole.

\subsection{Physical node lifetime and community membership} 
We define the \emph{lifetime of a physical node} as the total number of distinct timestamps where it appears in the temporal graph --- it is fundamentally a node property. 
Fig.~\ref{fig:node_lifetime_dist} represents the distribution of node lifetimes, which is similar for ACL and Blogs (it is also similar whether node lifetime is based on link targets only, or on link targets and origins): almost 40\% of nodes just appear  for a single timestamp; the distribution is heterogeneous with many nodes exhibiting a very weak activity.


Physical nodes may belong to different number of communities in their lifetime. Fig.~\ref{fig:comm_mem_dist} shows the fraction of physical nodes belonging to a given number of temporal communities, or \emph{community membership}. While most nodes belong to a single community throughout their lifetime, a significant number of nodes are members of various communities as well; following, again, a pattern typical of an heterogeneous probability law. There is admittedly a relationship between node lifetime and community membership, as a short-lived physical node is likely to belong to few communities. Fig.~\ref{fig:comm_mem_dist} certainly shows that in ACL, around two-thirds of authors belong to one community throughout their lifetime. It is unclear, however, whether all nodes belonging to few communities are short-lived: they may also constantly belong to a single community. To address this issue, we plot the correlation between node lifetime and community membership on Fig.~\ref{fig:n_comm_lifetime_assort}. There are indeed few committed nodes belonging to a single temporal community for a very long time (Blogs: maximum 151 timestamps, ACL: 23 years); but the landscape is globally mixed: while nodes appearing only once (trivially) belong to a single community, and few other nodes are members of multiple communities in their lifetime.



\subsection{Multiplicity and change in community membership}

\begin{figure*}
\begin{center}
\subfigure[Community multiplicity $C_M$ distribution]{\label{fig:comm_change_freq}\includegraphics[width=.45\linewidth]{\figs/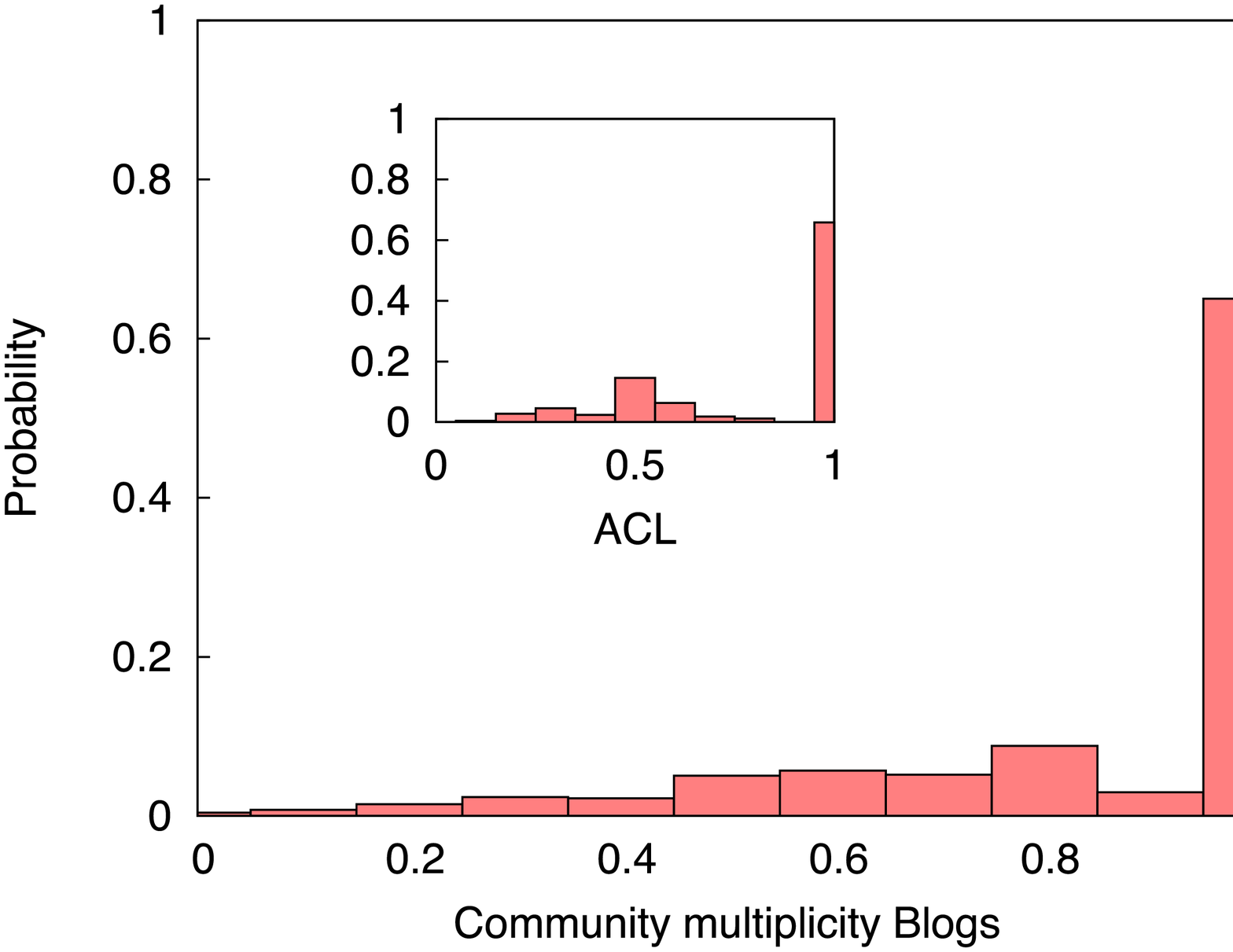}}
\hspace{0.5cm}
\subfigure[Community toggle $C_T$ distribution]{\label{fig:comm_togg_freq}\includegraphics[width=.45\linewidth]{\figs/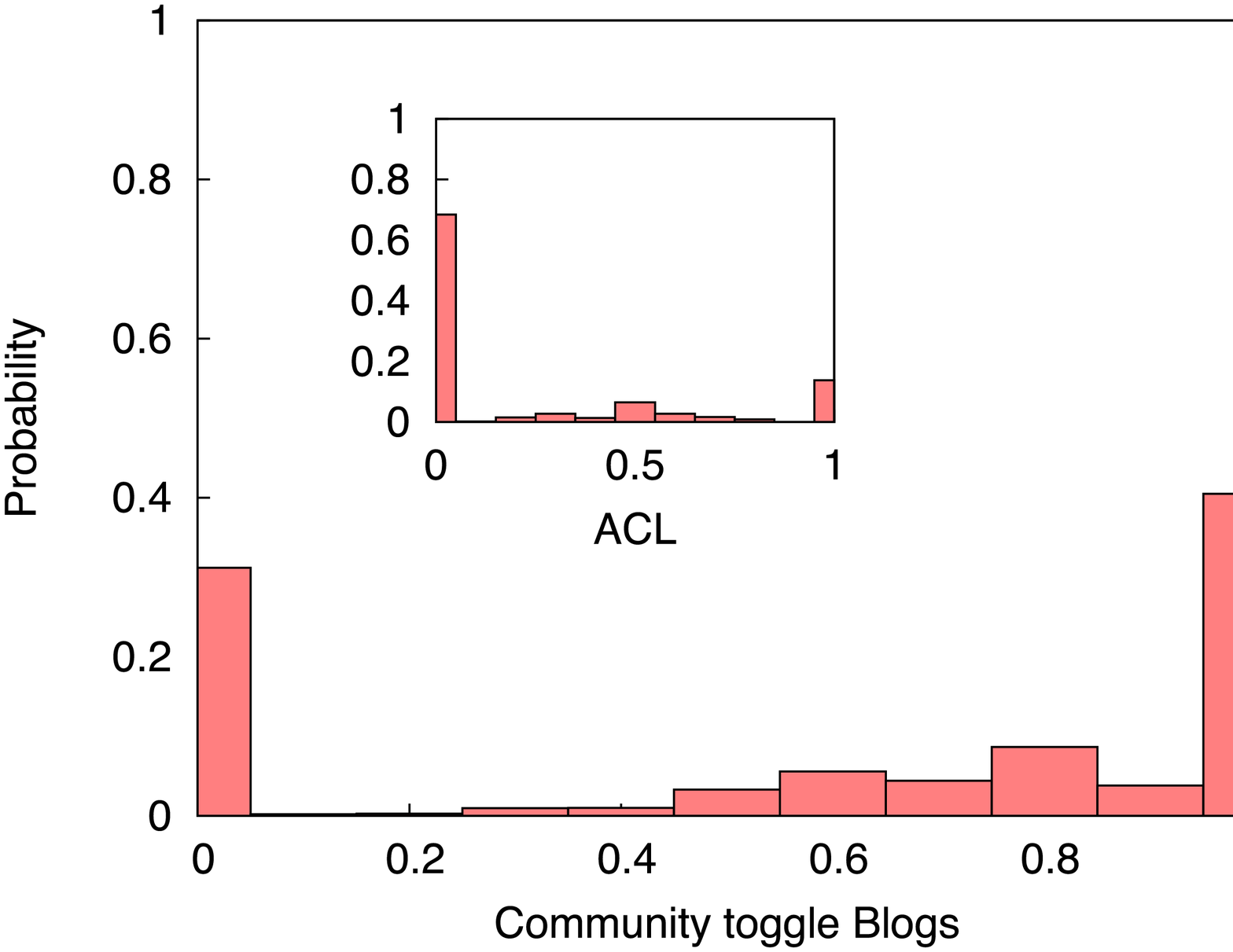}}
\end{center}
\vspace{-0.2in}
\caption{Binned distributions of community multiplicity or toggle indices for all nodes. Main figures correspond to ``Blogs'' data, insets to ``ACL''.}
\end{figure*}

More precisely, we examine the distribution of the multiplicity of community membership $C_M$ defined as the ratio of the node membership over its lifetime ---  see Fig.~\ref{fig:comm_change_freq}. This value equals $1$ when a node belongs to as many different communities as the number of timesteps where it is active, and is closer to zero when it is involved in a smaller number of communities. About 60\% of nodes have a $C_M$ of $1$, while around 10-20\% of nodes have a $C_M \in \left] 0.45;0.55 \right] $. 

Where nodes with smaller $C_M$ certainly migrate seldom from a temporal community to another, nodes with higher $C_M$ values might as well be heavily migrating or just short-lived nodes. To concretely appraise these migration or toggling effects, we finally define the ``community toggle'' $C_T$ of a node as the ratio between the number of times it jumps from a temporal community to another in the next active timestep, over the maximum number of toggles it can do, which is its lifetime minus one (when lifetime is 1 there is no possible toggle so $C_T=0$ by definition). We plot the distribution of $C_T$ on Fig.~\ref{fig:comm_togg_freq}. We can observe that it is generally bimodal: there are therefore two types of nodes.  Nodes constantly belonging to a community (be it with a short or long lifetime), or nodes belonging to a variety of communities. We can observe that $30\%$ of bloggers exhibit a very low toggle rate (between 0.0-0.1, mostly 0), whereas a similar proportion of nodes has heavily changing temporal community memberships -- perhaps a typical feature of blogspace. In ACL however, authors exhibit mostly stable behavior with a very low toggle rate for $70\%$.

This last argument underlines a possible issue in usual partitioning algorithms, with respect to detecting whether a node is constantly central in a given community, or whether it is sporadically marginal, peripheral or even a bumblebee for a series of communities. Node toggling statistics, possibly coupled with other dimensions such as degree, could be fruitfully used to discriminate such kinds of nodes. ``Backbone nodes'', for instance, are active, often self-citing and have a very low toggling rate, as is the case for example of \url{developpez.net} (rate = 0.024): a forum-like blog which is one of the most important French-speaking website in programming. On the other hand, \url{ump-senat.fr} has a high toggling rate (rate = 0.955): this is the blog of the right-wing group at the French Senate, very active yet on a variety topics and, more importantly, never really central on any of them.

\section
{\TB{Extension of the methodology}\label{sec:discussion}}
\BM{In this section, we suggest several different extensions of our methodology. First we show that our algorithm is also applicable beyond the 
citation dataset. Next, we propose a possible improvement of our algorithm through a complementary \emph{community repairing process}.}
\subsection{\TB{Scope: other types of datasets and networks}} We admittedly focused on citation datasets, where link extremities generally correspond to distinct times (at $t$, B cites an instance of A at $t'<t$). We now suggest that other types of non-strictly citation datasets can be appraised.

First, we may use all co-appearance dynamic networks, which are usually well represented by bipartite graphs. In effect, whenever some nodes of one side of the bipartite graph are linked to one node of the other side, at possibly different timesteps, it is possible to link the first-side nodes together. Many monopartite networks are actually projections of bipartite networks \cite{guil:bipa}, where $A$ and $B$ get connected because they were jointly connected to a common $r$; yet, traditional projections overlook the fact that $A$ and $B$ may have been linked to $r$ at different times. 
An example would be peer-to-peer exchange networks: if $A$ makes a request for resource \emph{r} at $t_A$ and then $B$ makes a similar request at $t_B > t_A$, we can create a timestamped link from $(A,t_A)$ to $(B,t_B)$. The corresponding  temporal graph captures the shared interest/involvement of some nodes across various timesteps.

More generally, the analysis we made on diachronic datasets draws benefits from the fact that interacting agents are associated with a specific timestamp. Our approach could nonetheless be used on a dataset which can be expressed as a sequence of interactions such as: $\left((v_i,\alpha_i),(v_j,\alpha_j)\right)$
where $\alpha_i\in\{\alpha_k\}_{k \in K}$ refers to a property of node $v_i$ rather than a timestamp. A node may for instance be  characterized by its location in space, thus enabling geographically-based communities: this in turn could be  useful in order to describe communities embedded in their geographical environment. Communication networks (phone calls, instant messaging, emails) could provide us with such data.

\subsection{\TB{Reconstruction: using the physical node information}\label{sec:merge}}

\begin{figure*}
\centering
\newcommand{\lmerge}{.138\linewidth}
\footnotesize\begin{tabular}{cccccc}
\includegraphics[width=\lmerge]{\figs/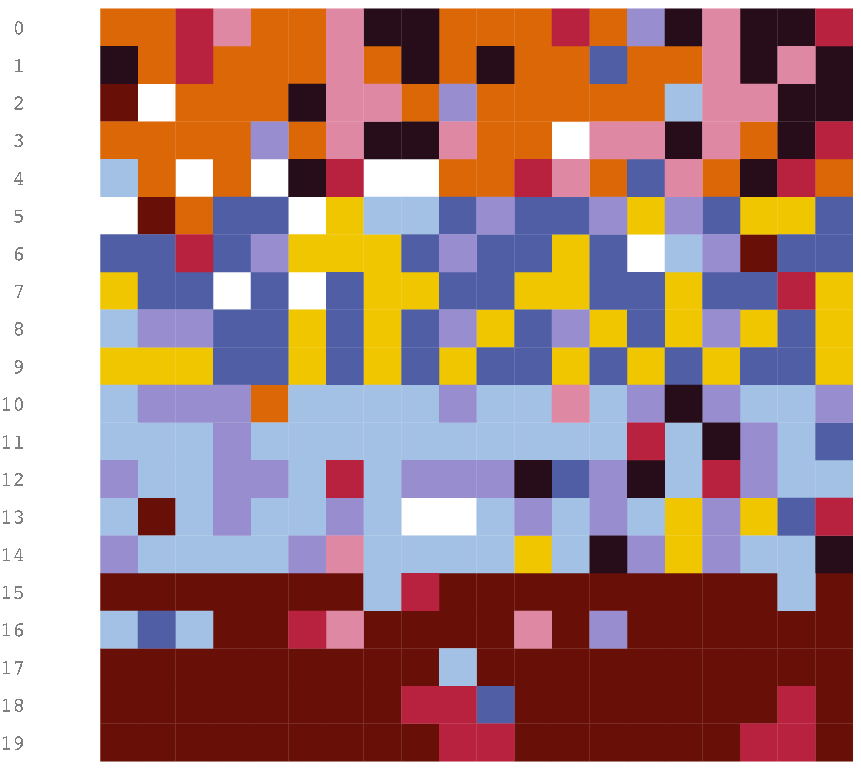}&
\includegraphics[width=\lmerge]{\figs/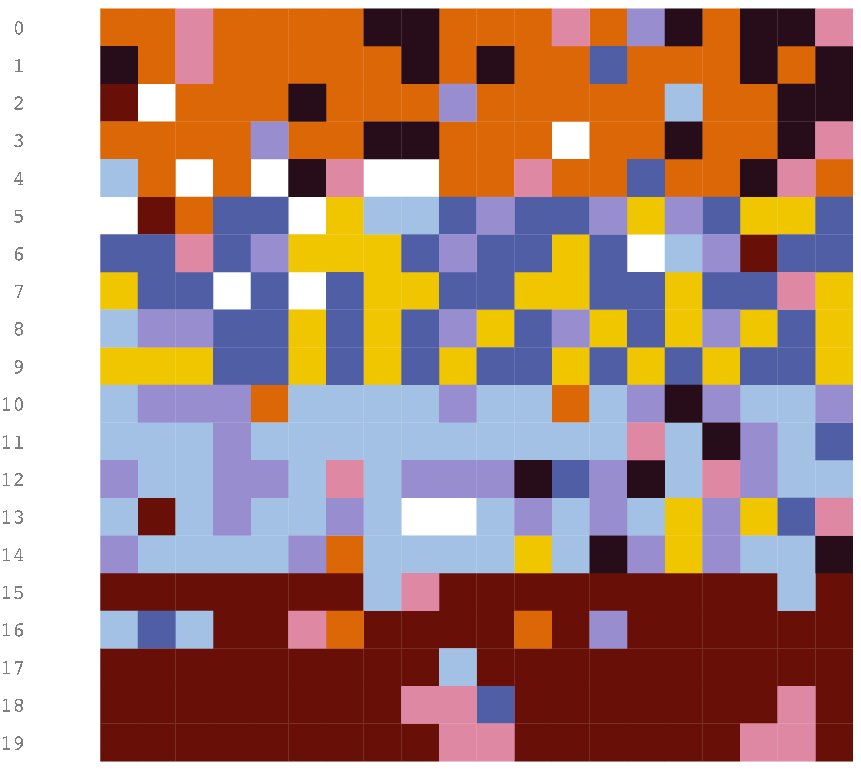}&
\includegraphics[width=\lmerge]{\figs/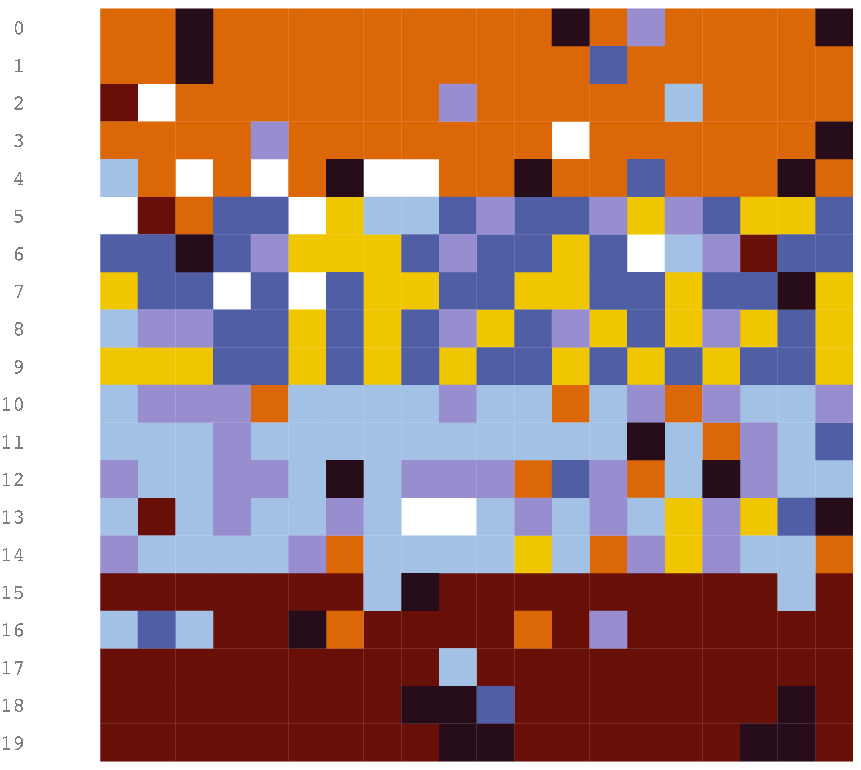}&
\includegraphics[width=\lmerge]{\figs/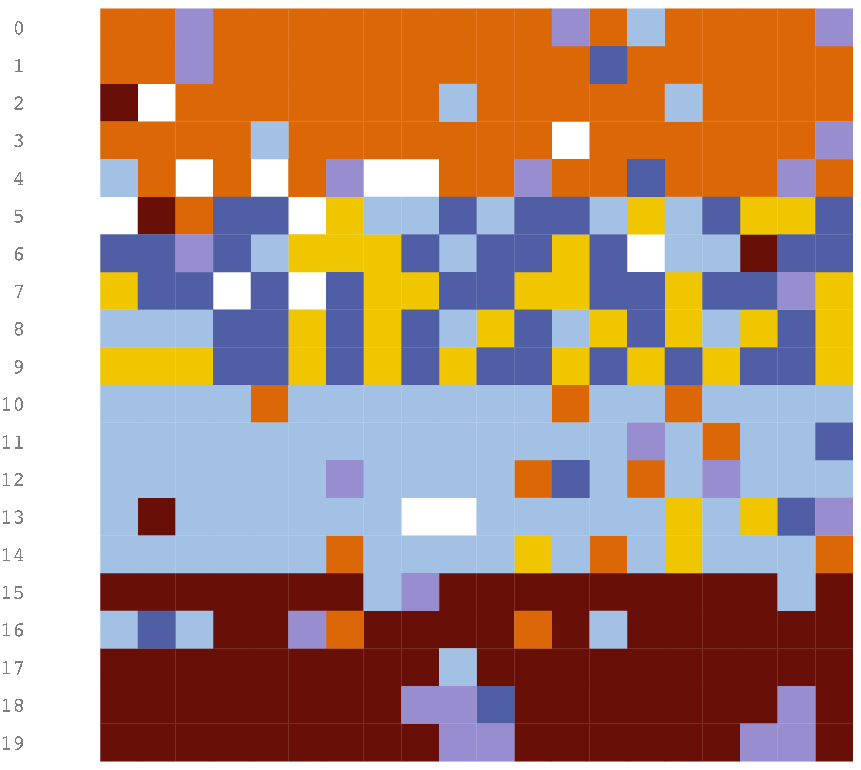}&
\includegraphics[width=\lmerge]{\figs/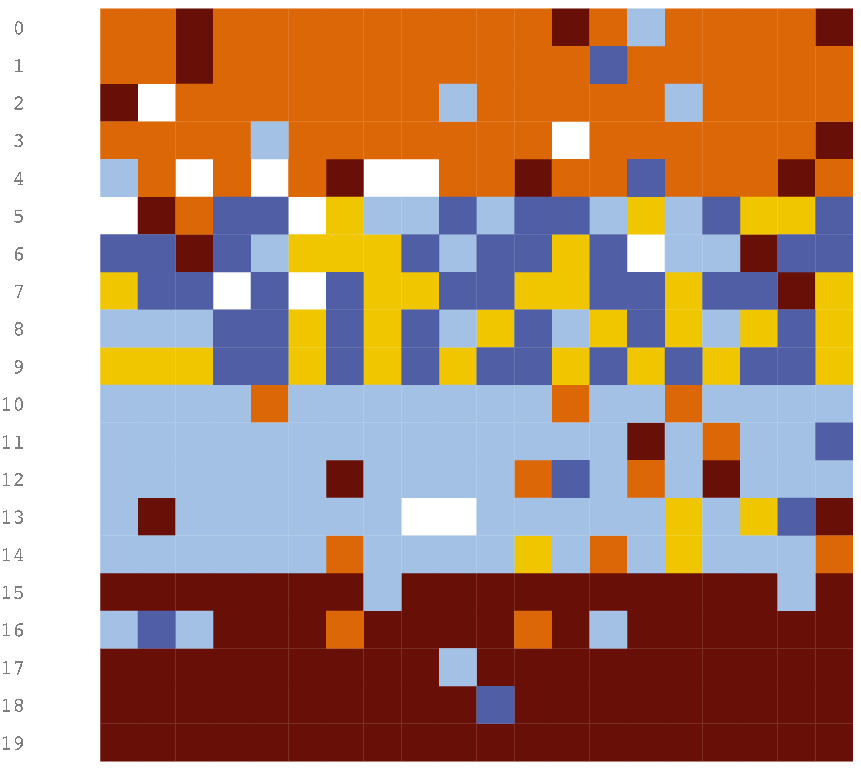}&
\includegraphics[width=\lmerge]{\figs/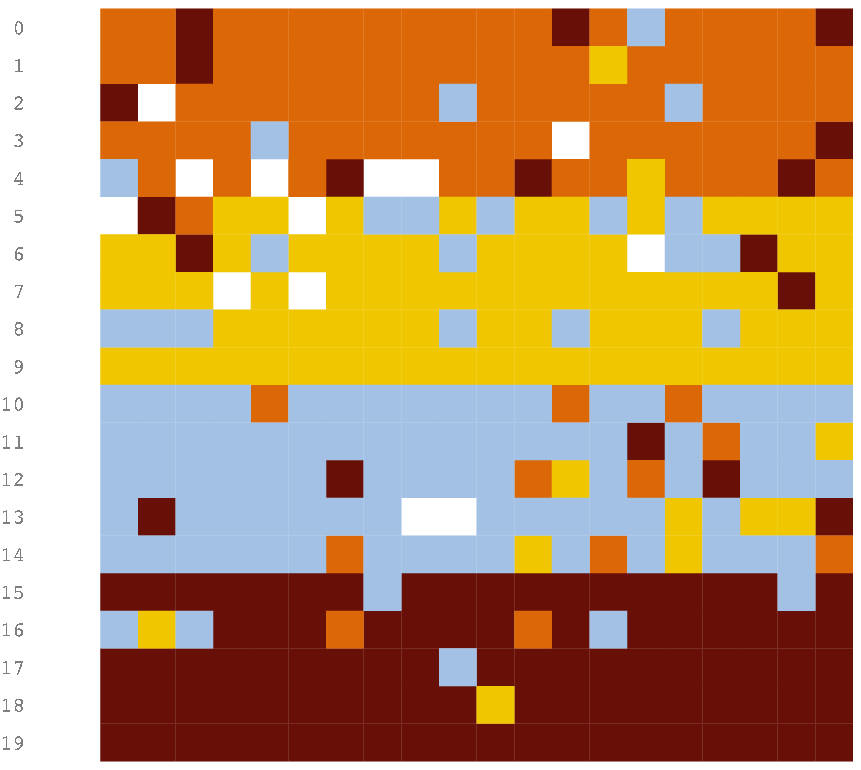}\\
Original&Step 1&Step 2&Step 3&Step 4&Step 5
\end{tabular}
\caption{\TB{Example of a portion of merging process based on $NA$ optimization, leading to physically more cohesive communities, from ``primary'' temporal communities.}\label{fig:mergexample}}
\end{figure*}

\TB{
Our methodology primarily aims at identifying  cohesive cross-temporal interaction patterns in the underlying diachronic data set. 
This approach focuses on structural properties, and cluster detection does not use contextual data yet, such as physical node information, or semantic data. It can thus be further improved by introducing an additional ``context-oriented'' step which can suitably complement the ``structure-oriented'' logic of our algorithm.}
\TB{This step would consist in reshaping the detected communities to optimize their contextual relevance. A straightforward example of \emph{context-aware community repairing} process can be described as the merging of different temporally-cohesive communities into a single physically-cohesive cluster. More precisely, temporal nodes duplicated from the same physical node supposedly share common features or interests \cite{jdidia}, and this information can suitably be used in a second step to restore or strengthen physical consistency in temporal communities detected by the proposed algorithm. For instance, the synthetic datasets defined in Section~\ref{sec:exp} rely on the assumption that physical nodes remain in a same \textit{a priori} community throughout their lifetime. We may thus reasonably expect that merging temporal communities consisting of similar physical nodes would generate communities closer to the \textit{a priori} assignment.  An illustrative toy example is presented on Fig.~\ref{fig:mergexample}, where such merging process is being essentially guided by an incremental optimization of physical cohesiveness through node activity $NA$ (it is however outside of the scope of this paper to dig further into such approach).}
\TB{In this framework, dynamic communities as we detect them with our method, could be combined and merged in order to build larger yet physically consistent dynamic communities, \hbox{i.e.} where dynamic communities based on similar physical nodes are eventually grouped. This notion of context may also be extended to more context-dependent features, such as semantic properties.}

\section{Conclusion}

Many existing dynamic community detection methods are based on successive snapshots and are therefore essentially longitudinal rather than dynamic.  By contrast, assuming that network communities fundamentally correspond to recurrent interaction over time, we proposed a methodology which does not alter the original temporal information of dynamic network data. In other words, it avoids a step where the original interaction data is aggregated into various network snapshots, or where connections are modified to accommodate relationships between nodes at various timesteps.  This requirement primarily led us to focus
on a specific type of network data where link extremities naturally correspond to distinct moments, as is typically the case in \emph{citation networks}.  We suggest however that the approach could be applied on several other widespread types of networks --- such as bipartite graphs, or more precisely co-use/co-citation/co-mention/co-appearance networks.

We introduced intrinsically dynamic metrics to qualify temporal community structure: we could therefore single out communities based on specific patterns of repeated inter-temporal interactions, internal hierarchization, and discuss in particular how and to what extent nodes switch from a community to another across time. More broadly, using macro-level visualizations of the whole temporal community structure, we could demonstrate that various empirical contexts exhibit distinct temporal community profiles, additionally emphasizing the fact that those various contexts may call for distinct `community' definitions and detection criteria. 

\bigskip
{\small\paragraph*{Acknowledgements}
We would like to thank Thomas Aynaud for his useful comments on the state-of-the-art.
This work has been partially supported by the French National Agency of Research (ANR)
through grant ``Webfluence'' \#ANR-08-SYSC-009 and by the Future and Emerging Technologies programme FP7-COSI-ICT of the European Commission through project QLectives (grant no.: \texttt{231200}).
}

\end{document}